\documentclass{CUP-JNL-DTM}%

\usepackage{graphicx}
\usepackage{subcaption}
\usepackage{caption}
\usepackage{multicol,multirow}
\usepackage[figuresright]{rotating}
\usepackage{array}
\usepackage{geometry}

\usepackage{tabularx}

\usepackage{appendix}
\usepackage[sort&compress,numbers]{natbib}
\usepackage{ifpdf}
\usepackage[T1]{fontenc}
\usepackage[utf8]{inputenc}
\usepackage{textcomp}
\usepackage{xcolor}
\definecolor{latent-purple}{RGB}{89,85,215}
\usepackage{nicematrix}
\usepackage{seqsplit}
\usepackage{tikz}

\usepackage{amsthm}
\usepackage{amsmath,amssymb,amsfonts}
\usepackage{newtxtext}
\usepackage{newtxmath}

\theoremstyle{definition}

\numberwithin{equation}{section}
\usepackage{diagbox}
\usepackage{float}

\usepackage[colorlinks,allcolors=latent-purple]{hyperref}
\usepackage{cleveref}
\crefname{appendix}{App.}{Apps.}
\crefname{section}{Sec.}{Secs.}
\crefname{subsection}{Sec.}{Secs.}
\crefname{subsubsection}{Sec.}{Secs.}
\crefname{figure}{Fig.}{Figs.}
\crefname{table}{Tab.}{Tabs.}
\crefname{equation}{Eq.}{Eqs.}
\crefname{algorithm}{Alg.}{Algs.}
\crefname{example}{Ex.}{Exs.}
\crefname{page}{p.}{pp.}
\crefname{line}{l.}{ll.}

\usepackage{siunitx}
\DeclareSIUnit\angstrom{\text{Å}}
\DeclareSIUnit\rpm{rpm}

\usepackage[acronym]{glossaries}
\makeglossaries
\glsdisablehyper

\newacronym{spr}{SPR}{surface plasmon resonance}
\newacronym[
    plural={T-cell engagers},
    firstplural={T-cell engagers (TCEs)}
]{TCE}{TCE}{T-cell engager}
\newacronym{igg}{IgG}{immunoglobulin G}
\newacronym{scfv}{scFv}{single-chain variable fragment}
\newacronym{vhh}{VHH}{variable heavy-chain-only}
\newacronym{fv}{Fv}{fragment variable}
\newacronym{cdr}{CDR}{Complementarity-Determining Region}
\newacronym{pdb}{PDB}{Protein Data Bank}
\newacronym{sabdab}{SAbDab}{The Structural Antibody Database}
\newacronym{bli}{BLI}{bio-layer interferometry}
\newacronym{pbmc}{PBMC}{peripheral blood mononuclear cells}
\newacronym{phd2}{PHD2}{prolyl hydroxylase domain-containing protein 2}

\usepackage{fancyhdr}
\fancyheadoffset{0pt}
\newcommand{\papertitle}{\raggedright Drug-like antibodies with low immunogenicity in human panels designed with Latent-X2}

\usepackage[version=4]{mhchem}

\newcommand{\themodel}{{Latent-X2}}
\newcommand{\thepreviousmodel}{{Latent-X1}}

\newcommand{\catno}{cat. no. }

\newcommand{\invivo}{\textit{in vivo}}
\newcommand{\exvivo}{\textit{ex vivo}}
\newcommand{\denovo}{\textit{de novo}}
\newcommand{\Denovo}{\textit{De novo}}
\newcommand{\insilico}{\textit{in silico}}
\newcommand{\Insilico}{\textit{In silico}}

\newcommand{\kd}{\ensuremath{\mathrm{K}_{\mathrm{D}}}}
\newcommand{\kon}{\ensuremath{\mathrm{k}_{\mathrm{on}}}}
\newcommand{\koff}{\ensuremath{\mathrm{k}_{\mathrm{off}}}}

\newcommand{\rampicon}{%
  \tikz[baseline=-0.2ex] 
    \fill (0,0) -- (1em,0) -- (1em,0.4em) -- cycle;
}

\newcommand{\kras}{K-Ras(G12D)}

\newcommand{\subfigref}[2]{Fig.~\hyperref[#1]{\ref*{#1}#2}}

\newcommand{\customcaption}[1]{%
  \refstepcounter{figure}%
  \caption*{\textbf{Fig.~\thefigure{}~|}~#1}%
}

\newcommand{\customtablecaption}[1]{%
  \refstepcounter{table}%
  \caption*{\textbf{Tab.~\thetable{}~|}~#1}%
}

\newcommand{\ArtType}{}
\makeatletter
\newcommand{\@opjournalheader}{}
\makeatother
\newcommand{\headeright}{}

\begin{document}
\setcounter{figure}{0}
\setcounter{table}{0}

\begin{Frontmatter}
\title[Article Title]{%
    \papertitle
}

\author[]{Latent Labs Team}
\address[]{\orgaddress{\city{London, UK \& San Francisco, USA}\\
           \orgaddress{16 December 2025}} \\ 
\begin{center}
     \makebox[\textwidth][c]{\includegraphics[width=1.0\textwidth]{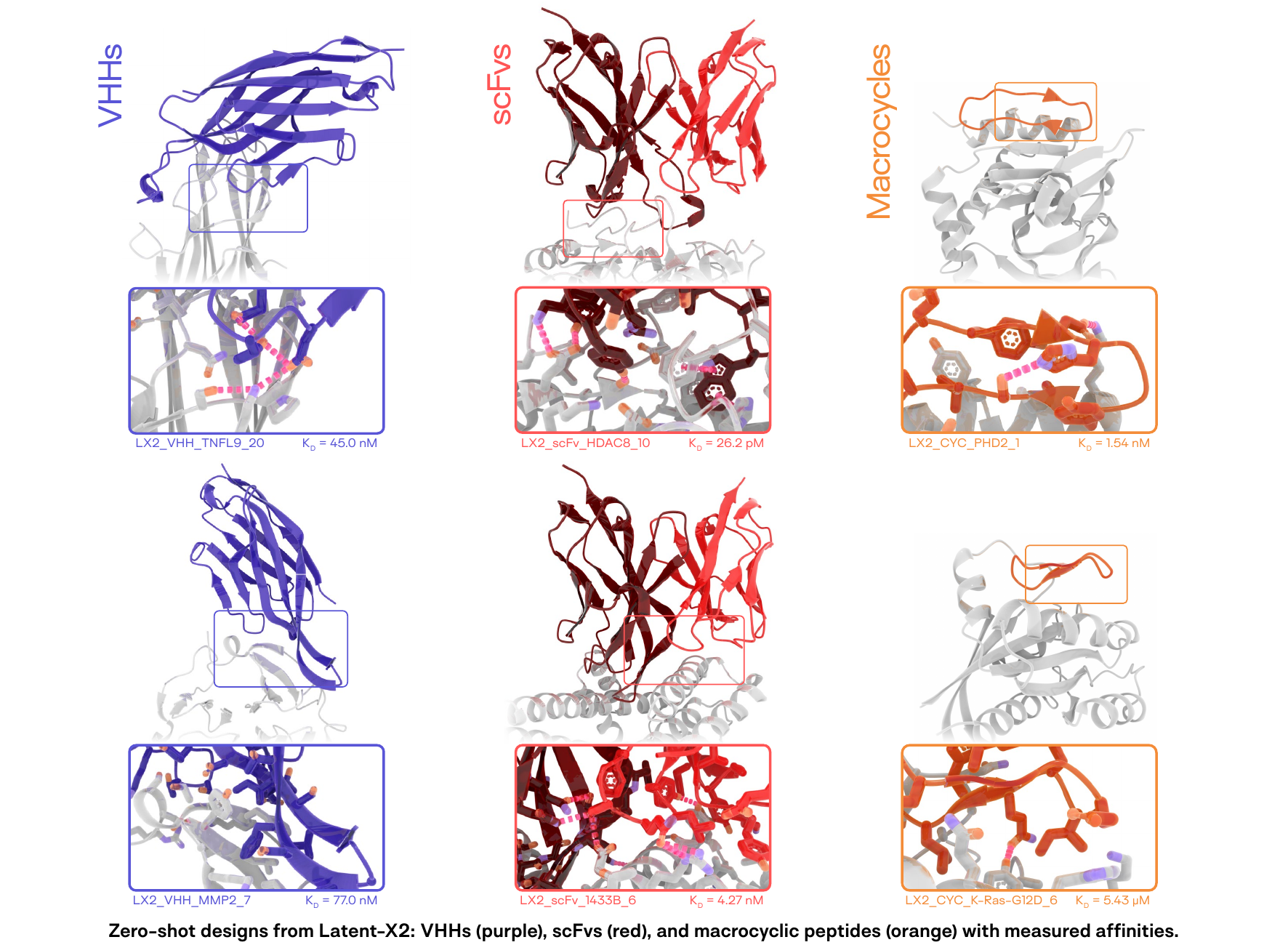}}
\end{center}
\vspace{-1cm}
}

\abstract{
Drug discovery has long sought computational systems capable of designing drug-like molecules directly: developable and non-immunogenic from the start. Here we introduce \themodel{}, a frontier generative model that achieves this goal through zero-shot design of antibodies with strong binding affinities, drug-like properties, and, for the first time for any \denovo{} generated antibody, confirmed low immunogenicity in human donor panels. \themodel{} is an all-atom model conditioned on target structure, epitope specification, and optional antibody framework, jointly generating sequences and structures while modelling the bound complex. Testing only 4 to 24 designs per target in each modality, we successfully generated VHH and scFv antibodies against 9 of 18 evaluated targets, achieving a \SI{50}{\percent} target-level success rate with picomolar to nanomolar binding affinities. Designed molecules exhibit developability profiles that match or exceed those of approved antibody therapeutics, including expression yield, aggregation propensity, polyreactivity, hydrophobicity, and thermal stability, without optimization, filtering, or selection. In the first immunogenicity assessment of any AI-generated antibody, representative \denovo{} VHH binders targeting TNFL9 exhibit both potent target engagement and low immunogenicity across T-cell proliferation and cytokine release assays. The model generalizes beyond antibodies: against K-Ras, long considered undruggable, we generated macrocyclic peptide binders competitive with trillion-scale mRNA display screens. These properties emerge directly from the model, demonstrating the therapeutic viability of zero-shot molecular design, now available without AI infrastructure or coding expertise at \href{https://platform.latentlabs.com}{\texttt{\color{latent-purple}https://platform.latentlabs.com}}.
}
\end{Frontmatter}

\section[Introduction]{Introduction}

Drug discovery has long pursued computational systems that directly design lead candidates, generating molecules that clear developability hurdles and do not trigger immune responses, while avoiding lengthy optimization. We report that this goal is now within reach. \themodel{} can generate antibodies that bind with picomolar to nanomolar affinity, exhibit drug-like developability, and, in the first demonstration for any AI-generated antibody, show low immunogenicity in human donor panels, all without post-generation optimization.

Current approaches are costly both in development and clinical failure. Molecules often fail not because they lack binding, but because binding alone is insufficient when clinical success demands developability and low immunogenicity. Optimization to address these liabilities frequently fails or produces zero-sum trade-offs, with some properties improving at the expense of others \cite{Jain2017biophysical}. This challenge consumes large budgets per clinical program and leaves critical targets difficult or impossible to address \cite{carter2024immunogenicity}.

Recent AI methods demonstrated the tractability of zero-shot binder generation \cite{watson2023novo, zambaldi2024novo, pacesa2024bindcraft, chai2025chai, team2025latent, mille2025efficient, stark2025boltzgen}, and concurrent work has shown that AI-generated antibodies can clear key developability thresholds \cite{bio2025novo, chai2025drug}. Building on the success of \thepreviousmodel{} \cite{team2025latent} for mini-binder and macrocycle design, \themodel{} advances further: the first low immunogenicity demonstration for any AI-generated antibody, and generalization across antibodies, macrocyclic peptides, and mini-binders in a single system. Computational benchmarking shows improved success rates over \thepreviousmodel{} for mini-binders and macrocycles, see \cref{app:insilico_hit_rates}.

\begin{figure}[h!]
    \centering
    \includegraphics[width=0.7\textwidth]{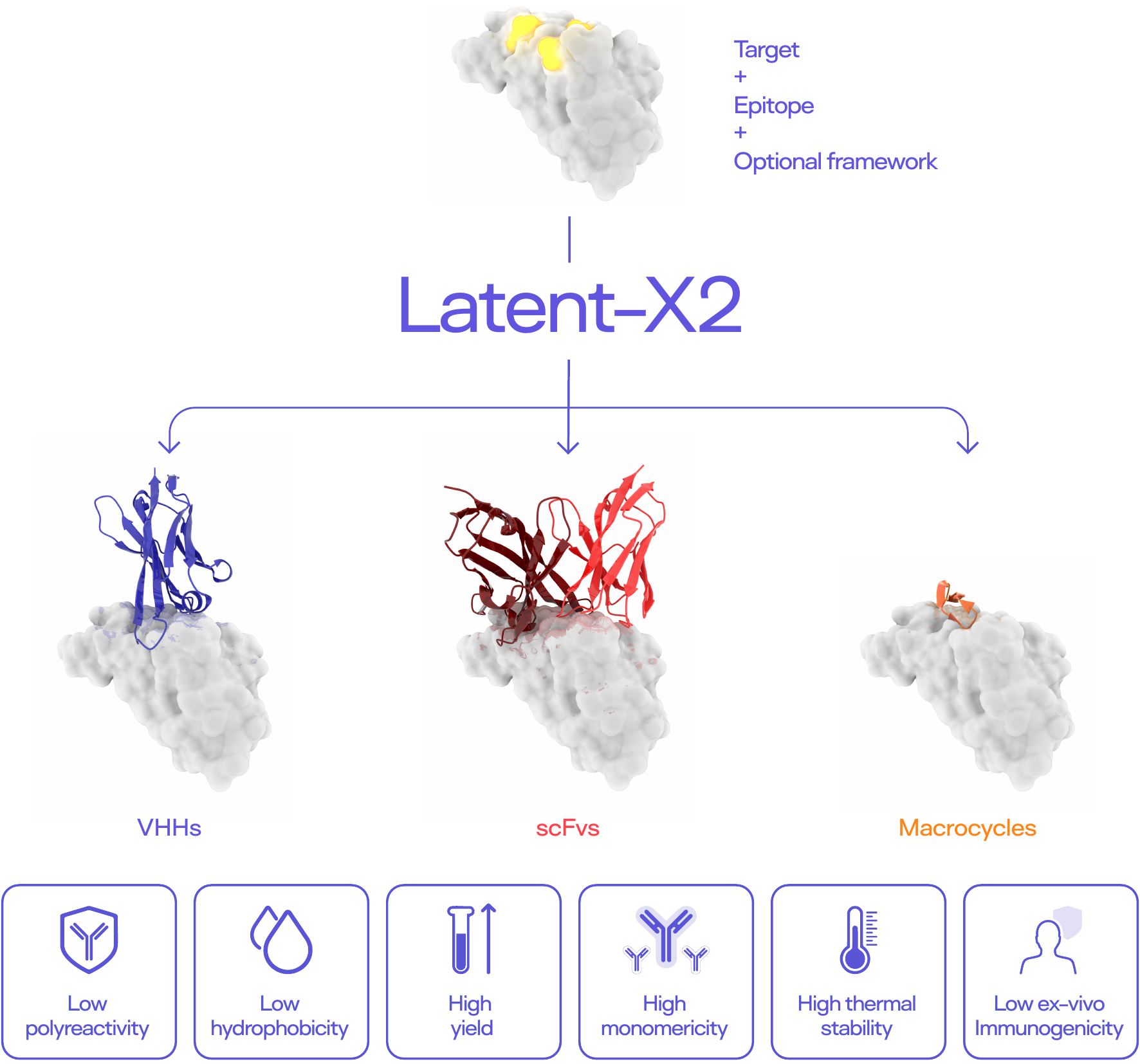}
    \vspace{15pt}
    \customcaption{\textbf{\themodel{} generates drug-like binders from target structure and epitope specification.} The multi-modal prompt comprises target structure (grey) with epitope hotspots (yellow) and an optional antibody framework. The model outputs binder sequence and structure in complex with the target across multiple modalities, including VHH, scFv and macrocyclic peptides. Designs exhibit favourable developability and low immunogenicity.}
    \label{fig:workflow}
\end{figure}

To achieve this, the model jointly generates all-atom structures and sequences conditioned on multi-modal prompts that comprise target structure, epitope specification, and optional antibody framework. An overview of this workflow is given in \cref{fig:workflow}. By directly modelling non-covalent interactions in the bound complex, it produces binders across \gls{vhh} and \gls{scfv} antibodies, as well as macrocyclic peptides, without task-specific fine-tuning.

We evaluated \themodel{} against 18 targets, testing only 4 to 24 antibody designs in each modality per target, achieving a \SI{50}{\percent} target-level success rate with high binding affinities. Separately, for \kras{} and \gls{phd2}, two challenging protein targets with relevance to oncology and hypoxia biology, we tested 10 macrocyclic peptides designed with \themodel{} that matched or exceeded hits from state-of-the-art mRNA display screens \cite{goto2021rapid}.

Drug-like properties emerged directly from the model. Designed antibodies exhibit high expression yield, favorable biophysical characteristics, and low polyreactivity, with developability profiles meeting or surpassing approved therapeutic benchmarks.

Notably, we present immunogenicity data for AI-generated antibodies. \Denovo{} VHH binders targeting TNFL9 were assessed across a ten-donor human panel, confirming both potent target engagement and low immunogenicity \exvivo{}. While animal studies and clinical trials remain ahead, these results demonstrate that AI-generated molecules can now clear preclinical hurdles that previously required lengthy optimization.

\textbf{Our main contributions are:}
\begin{enumerate}
\item First \denovo{} designed antibodies with demonstrated low immunogenicity in human donor panels.
\item Zero-shot design with \SI{50}{\percent} target-level success across 18 targets, testing only 4 to 24 designs per modality.
\item Picomolar to nanomolar binding affinities with developability matching or exceeding approved therapeutics.
\item A single architecture spanning VHHs, scFvs, macrocyclic peptides, and mini-binders.
\item Macrocyclic peptide binders matching or exceeding trillion-scale mRNA display screens.
\end{enumerate}

Below we introduce the workflow and provide detailed wet lab validation. Apply for access to \themodel{} on \href{https://platform.latentlabs.com}{\texttt{\color{latent-purple}https://platform.latentlabs.com}}.

\section[Results]{Results}
\label{sec:results}

Across antibody and macrocycle design, we demonstrated that \themodel{} produces high-affinity binders in a zero-shot regime. Testing only 4 to 24 designs in each modality-target pair, we experimentally characterized designs against structurally diverse targets. For antibodies, the model generates potent \glspl{vhh} and \glspl{scfv} with drug-like biophysical, expression, and polyreactivity profiles from the first generation. For the target TNFL9, representative VHH binders showed no detectable immunogenic response in \exvivo{} T-cell proliferation and cytokine release assays. For macrocyclic peptides, \themodel{} designs high-affinity binders against \kras{} and \gls{phd2} with performance competitive with state-of-the-art experimental discovery platforms, achieving single-digit nanomolar affinity against \gls{phd2}. Together, these results establish that a single all-atom generative system can achieve drug-like antibody design and high-affinity macrocycle generation across multiple target profiles and critical preclinical development hurdles.

\subsection{High-affinity VHH and scFv binders with drug-like developability}

\begin{figure}[h!]
    \centering
    \includegraphics[width=0.99\textwidth]{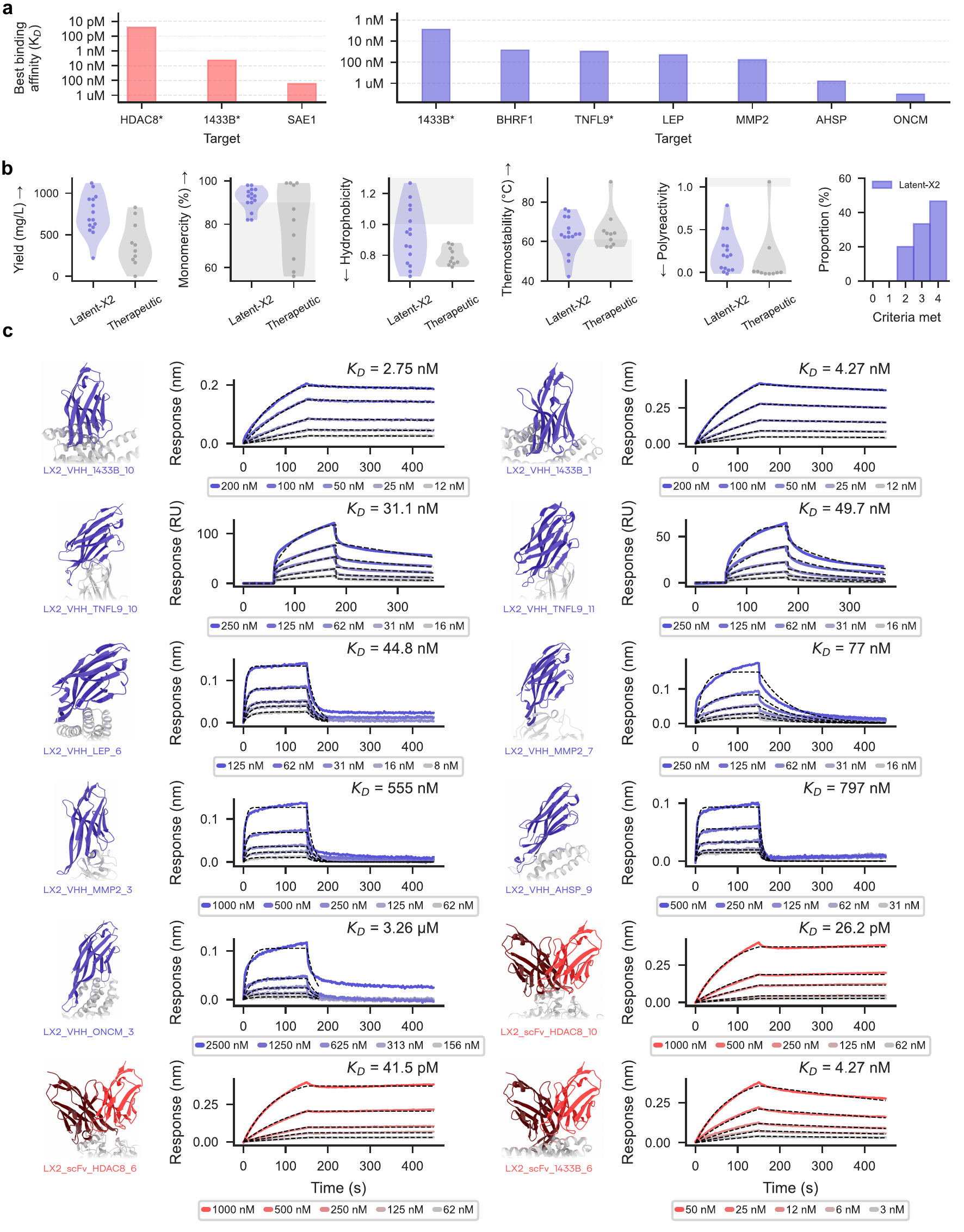}
    \customcaption{\textbf{High-affinity \gls{vhh} and \gls{scfv} binders with drug-like developability.} a) Best binding affinity (\kd) per target for scFvs (red) and VHHs (purple). Asterisks denote affinities influenced by avidity effects. b) Developability profiles compared to therapeutic benchmarks; grey shading indicates unfavorable regions. The rightmost panel shows the distribution of binders by number of developability criteria met. c) Designed structures with corresponding BLI or SPR response curves.}
    \label{fig:vhh_scfv_binders}
\end{figure}

To test generalization, we used \themodel{} to generate \gls{vhh} and \gls{scfv} antibody designs for 18 soluble protein targets. We selected this panel for its diverse biological functions, biophysical challenges, and therapeutic relevance. 

\themodel{} generated validated binders for 9 of 18 targets, evaluating no more than 24 designs per antibody modality per target in a single round of zero-shot design. In the best-performing experiments, up to \SI{25}{\percent} of designs produced confirmed binders. Successful binders spanned a range of antibody scaffolds and \gls{cdr} lengths. Binding affinities were measured by 5-point \gls{bli} or \gls{spr}, yielding affinities as strong as low picomolar. The highest-affinity binder for each target is shown in \subfigref{fig:vhh_scfv_binders}{a}, with representative structures and response curves in \subfigref{fig:vhh_scfv_binders}{c}. The strongest binder achieved a \kd{} of \SI{26.2}{pM} against HDAC8, a class I histone deacetylase and emerging oncology target \cite{chakrabarti2016targeting}. For each target, we provide a lab-validated \themodel{} generated sequence in \cref{tab:binder_table} to allow for external reproduction. We publish the corresponding designed structures on \href{https://platform.latentlabs.com}{\texttt{\color{black}https://platform.latentlabs.com}}, accessible without sign-in.

Beyond binding affinity, we subjected all candidates to comprehensive developability profiling: expression yield, monomericity, hydrophobicity, thermostability, and polyreactivity. These properties govern manufacturability, stability, pharmacokinetics, and immunogenicity risk \cite{Jain2017biophysical, Jain2023identifying}. The distribution of confirmed binders across each metric is shown in \subfigref{fig:vhh_scfv_binders}{b}. Assay details are provided in \cref{sec:methods}.

Across all targets for which binders were obtained, \SI{47}{\percent} matched or exceeded thresholds derived from therapeutic antibodies on all four metrics (monomericity, hydrophobicity, thermostability, polyreactivity), with \SI{80}{\percent} doing so on three of four, demonstrating that drug-like properties emerged directly from zero-shot design \subfigref{fig:vhh_scfv_binders}{b}.

Beyond the controls we used to establish thresholds, we assessed ten \gls{vhh} and \gls{scfv} domains derived from approved and clinical-stage therapeutics to contextualise the developability profiles of our designs. These comparison antibodies are detailed in \cref{sec:developability_benchmark}. Because these antibodies are not natively expressed as isolated \gls{vhh} or \gls{scfv}, we reconstructed their relevant domains as Fc-VHH and Fc-scFv fusion constructs. Across all metrics, distributions for \themodel{} designs were comparable to, or in some cases better than, those of the benchmark antibodies, further demonstrating that zero-shot designs achieve developability on par with therapeutic antibodies.

\subsection{\Denovo{} antibodies with no detectable immunogenicity in human donor panels}
\label{sec:immunogenicity_results}

\begin{figure}[h!]
    \centering
    \includegraphics[width=0.95\textwidth]{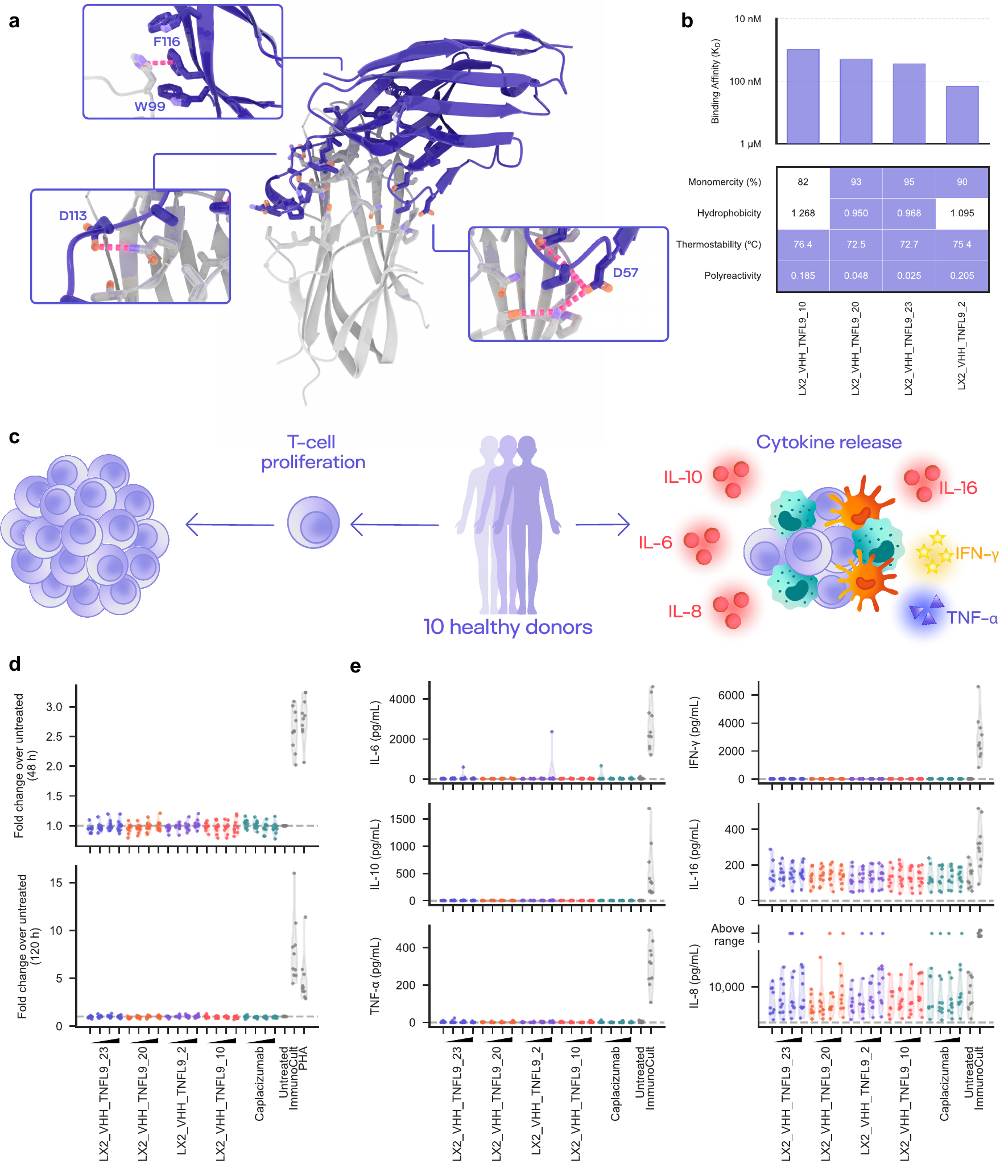}
        \customcaption{\textbf{\themodel{} \gls{vhh} binders to TNFL9 with drug-like developability and low immunogenicity.} a) Designed \gls{vhh} (purple) bound to TNFL9 (grey), with epitope interactions (pink) in close-up. Mutating any of F116, W99 or D113 to alanine eliminated binding. b) Binding affinities and developability metrics for four confirmed binders; in the developability heatmap, purple indicates metrics meeting criteria, while white indicates those that do not. c) \textit{Ex vivo} immunogenicity workflow across 10 healthy donors. d) T-cell proliferation at 48 and 120 hours; untreated control is shown alongside positive controls (ImmunoCult and PHA), with no increase observed for designed binders or caplacizumab (approved VHH-based therapeutic). e) Cytokine release at 120 hours; untreated control and ImmunoCult positive control are shown, with no elevation detected for designed binders or caplacizumab.  IL-8 measurements occasionally exceeded the assay’s upper detection limit across conditions, including ImmunoCult, caplacizumab, and designed binders, and are shown as “above range” on the axis. For panels d and e, increasing concentrations are denoted by \rampicon\ corresponding to 1.11, 3.33, 10, and \SI{30}{\mu g/mL}. Each point represents an individual donor; two replicates were performed per condition, with the mean shown where applicable.}
    \label{fig:tnfl9}
\end{figure}

The highest-risk stage of drug development is clinical translation, where the majority of candidates fail \cite{hay2014clinical, carter2024immunogenicity}. A major contributor to late-stage attrition is immunogenicity, where a therapeutic candidate provokes an unwanted immune response in a patient. For antibody-format therapeutics, immunogenicity can lead to anti-drug antibody formation, which may neutralize efficacy, accelerate drug clearance, alter pharmacokinetics, or trigger serious immune reactions \cite{vaisman2020molecular}. Immunogenicity represents a critical liability in biologic drug design and has directly contributed to the failure of otherwise promising clinical candidates \cite{ridker2017cardiovascular}.

To determine whether designs from \themodel{} extend beyond binding hits toward viable lead candidates, TNFL9 was selected as a representative target for deeper experimental characterization. TNFL9 is an immunomodulatory target implicated in inflammatory and immune signaling pathways, making it a representative and challenging target for therapeutic antibody development \cite{wang2009immune}. 

From 24 designs against TNFL9, five VHH binders were identified, of which four were selected for further analysis of binding specificity and immunogenicity. The four VHH binders were  experimentally confirmed to be low nanomolar affinity. These binders were next evaluated for developability: three passed three of four developability criteria, and two met all binding and developability criteria, see \subfigref{fig:tnfl9}{b}. Binding specificity was assessed by alanine mutagenesis of key CDRH3 residues; independent substitutions at three positions abolished binding in all four VHHs, consistent with target recognition being driven by the designed epitope interactions, an example is shown in \subfigref{fig:tnfl9}{a}.

To assess whether drug-like properties emerge from zero-shot design, we evaluated immunogenicity risk using primary immune cells isolated from the blood of 10 healthy human donors, see \cref{tab:immuno_controls} for anonymized donor characteristics. These \exvivo{} cellular tests directly measure whether a candidate induces human T-cell proliferation or triggers cytokine release—two of the most common readouts of immunogenic response used in therapeutic development. In T-cell proliferation assays measured at 48 and 120 hours, no increase was observed for any of the four \glspl{vhh} relative to background across all donors and time-points, see \subfigref{fig:tnfl9}{d}. In cytokine release assays at 120 hours, no elevation was detected for any \gls{vhh} across all donors, see \subfigref{fig:tnfl9}{e}. The approved \gls{vhh}-based caplacizumab exhibited a comparable profile. For assay details, see \cref{sec:immunogenicity}.

Together, these results indicate that VHHs designed by \themodel{} display not only high-affinity binding but also emergent drug-like properties, including favorable developability profiles and absence of detectable immunogenic responses in human primary-cell assays. The convergence of binding, developability, and low immunogenicity within the same candidates supports progression beyond hit discovery toward lead candidate selection. To our knowledge, this is the first reported evaluation of a \denovo{}–designed antibody using human cellular immunogenicity assays, and no measurable immunogenic signal was detected.

\subsection{Macrocyclic peptides outperforming trillion-scale mRNA display}
Macrocyclic peptide discovery conventionally requires screening vast molecular libraries to identify initial hits. mRNA display has emerged as a leading platform, enabling screens exceeding $10^{12}$ compounds, and has yielded several late-stage therapeutics including enlicitide; Merck's oral PCSK9 inhibitor currently in phase 3 trials \cite{johns2023orally}.

We benchmarked \themodel{} against RaPID (random nonstandard peptides integrated discovery) \cite{goto2021rapid}, a widely adopted mRNA display platform that end-to-end requires months per target to achieve a validated hit. We compared \themodel{} against mRNA display on two oncology targets for which RaPID-derived macrocyclic campaigns have been reported: \gls{phd2} \cite{mcallister2018non, chowdhury2020use}, a central regulator of the hypoxic response, and \kras{} \cite{zhang2020gtp}, one of the most prevalent oncogenic K-Ras mutations. For both targets, the strongest reported macrocycles from the literature were resynthesized and their binding affinities measured under identical assay conditions. Ten \themodel{} designs per target were selected for synthesis and evaluated experimentally within four weeks.

\begin{figure}[h!]
    \centering
    \includegraphics[width=0.99\textwidth]{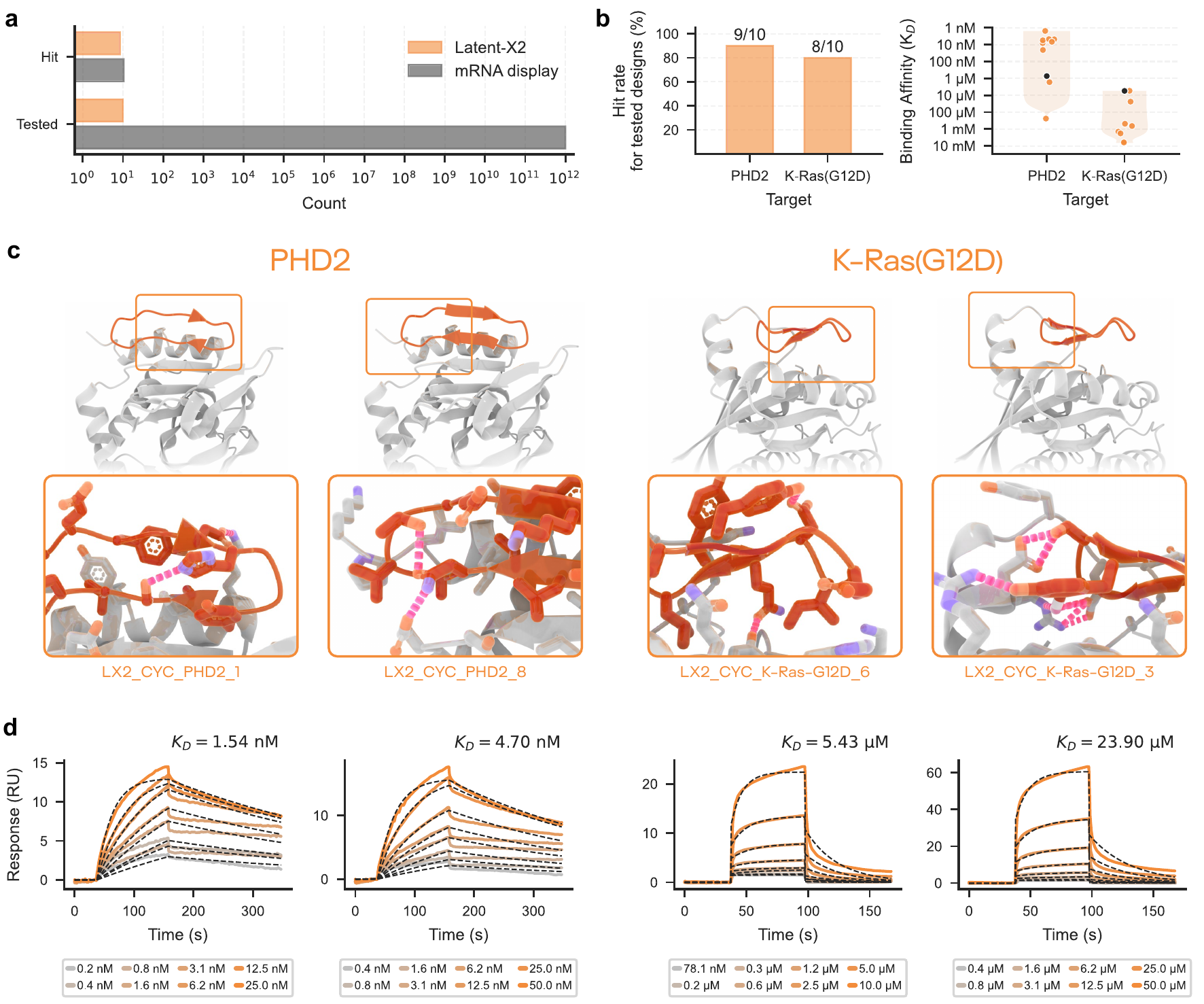}
    \customcaption{\textbf{Macrocyclic peptides outperforming trillion-scale mRNA display.} a) Comparison of experimental scale: \themodel{} tested 10 designs per target versus $>10^{12}$ compounds screened by mRNA display, with comparable or higher hit counts. b) Hit rates and binding affinities for \gls{phd2} and \kras{}. \themodel{} achieved 9/10 and 8/10 hit rates respectively. Black points indicate best replicated mRNA display affinities. c) Designed macrocycle structures (orange) bound to \gls{phd2} and \kras{} (grey), with epitope interactions (pink) shown in close-up. d) SPR response curves for top binders against each target.}
    \label{fig:macrocycle_figure}
\end{figure}

Using \themodel{}, high-affinity binders were obtained for both targets with substantially higher hit rates than RaPID, see \subfigref{fig:macrocycle_figure}{b,c}. For \gls{phd2}, 9 of 10 designs bound, with the strongest achieving \SI{1.54}{nM} affinity compared to \SI{729}{nM} for the best-reported RaPID hit. For \kras{}, 8 of 10 designs bound, with a best affinity of \SI{5.43}{\micro M}, matching the leading mRNA display hit (\SI{5.53}{\micro M}). The \themodel{} macrocycles exhibit diverse secondary structure, with \kras{} designs spanning $\beta$-sheet and $\alpha$-helical conformations. Successful designs ranged from 12–18 amino acids for \kras{} and 13–16 for \gls{phd2}, with the strongest \kras{} binder comprising only 12 residues, see \subfigref{fig:macrocycle_figure}{a}.

Prior mRNA display campaigns reported only 5 and 16 validated hits, respectively, from libraries exceeding one trillion members. \themodel{} achieved hit rates of \SIrange{80}{90}{\percent} from only 10 designs per target — a reduction in experimental search space of over 11 orders of magnitude while matching or exceeding binding affinity.

These results demonstrate, to our knowledge, the first instance of AI-driven peptide design matching or exceeding exhaustive experimental screening in head-to-head comparison. \themodel{} establishes a precision-first paradigm for macrocycle discovery: replacing brute-force library screening with generative design that yields lab-validated hits in weeks rather than months.

\section[Methods]{Methods}
\label{sec:methods}

\themodel{} is an all-atom frontier model that jointly designs binder structure and sequence, without refinement steps like resequencing of the generated structure. The model takes as input a prompt specifying: a target structure (which may span multiple chains), hotspot residues defining the epitope, and a binder modality. Supported modalities include antibodies, macrocyclic peptides, and mini-binders. For antibodies, a framework structure and desired \gls{cdr} lengths are additionally provided as input, allowing specification of different antibody formats like \gls{vhh} and \gls{scfv}. Given these inputs, the model generates a full binder–target complex with the binder engaging the specified epitope. Further details are provided in \cref{app:model_inputs}.

To select viable designs, we apply an \insilico{} filter based on structure prediction models \cite{abramson2024accurate, boitreaud2024chai, passaro2025boltz}. For antibody binders, we filter based on structure prediction confidence metrics and a self-consistency score comparing the \themodel{}-generated complex to the predicted complex structure via DockQ \cite{mirabello2024dockq}. For mini-binders and macrocycles, we use the scoring approach described in our previous work \cite{team2025latent}.

\Insilico{} pass rates for antibody designs vary by target, epitope, and framework: benchmarking on a held-out set of 200 target proteins with randomly sampled epitopes yielded an average pass rate of approximately \SI{1}{\percent}, with some configurations exceeding \SI{20}{\percent}. The distribution of pass rates is shown in \cref{fig:dist_plot_vhh}, see \cref{app:insilico_hit_rates}. Some configurations are difficult to achieve favorable computational metrics for, and can require $\sim\!10^3$ samples. For mini-binders and macrocycles, \themodel{} shows improved \insilico{} hit rates over \thepreviousmodel{} on the same test set, see \cref{fig:dist_plot_minibinders_and_macrocyles}. 

Owing to continued research progress, we used two variants of the same foundational model architecture.
\begin{itemize}
\item \textbf{\themodel{}:} used to generate binders for experimental validation.
\item \textbf{\themodel{}.1:} a slightly improved version used in studies of expected \insilico{} hit rates, see \cref{fig:dist_plot_vhh}. This is the \themodel{} model available through \href{https://platform.latentlabs.com}{\texttt{\color{latent-purple}https://platform.latentlabs.com}}.
\end{itemize}

\subsection{VHH and scFv binder design}
\label{sec:vhh_scfv_design_methods}
Epitopes were specified by defining hotspot residues on the target surface. Hotspot selection was informed by available structural data: for some targets, we used previously validated epitopes from prior work \cite{team2025latent}. For others, hotspots were derived from analysis of known binding interfaces or co-crystal structures. All targets are specified in \cref{tab:target_table}. For each target, we sampled one to four interface residues as hotspots.

We generated \gls{vhh} and \gls{scfv} designs using five frameworks each, see \cref{tab:vhh_frameworks} and \cref{tab:scfv_frameworks}, selected based on reported use in clinical trials or prior research. For scFvs, we computationally modelled the binder as an Fv and subsequently appended a flexible peptide linker to construct the full scFv. All CDR loops were generated \denovo{}, three for VHHs and six for scFvs, with no grafted or fixed regions. Given the dominant contribution of CDRH3 to antigen recognition \cite{tsuchiya2016diversity}, we varied CDRH3 length by up to $\pm\, 2$ residues relative to the native loop for each framework, while keeping all other CDR lengths fixed.

To ensure sequence novelty, we compared generated antibody sequences against \gls{sabdab} \cite{dunbar2014sabdab} by computing edit distance over concatenated \gls{cdr} loops; for \glspl{scfv}, computed individually for light and heavy chains. We performed sequence similarity searches using MMseqs2 \cite{steinegger2017mmseqs2}. The maximum number of reported matches was set to exceed the size of the target database, and E-value filtering was disabled, so that all matches passing MMseqs2’s prefiltering and alignment stages were retained (see \cref{app:novelty} for details). All designs advanced to experimental testing have a CDR edit distance of at least 11 to the closest match in \gls{sabdab}, with most exceeding 20. The closest confirmed binder has a CDR edit distance of 18 residues, indicating that successful designs are not a result of memorization. We performed an additional novelty check using BLASTP \cite{altschul1990basic} against the non-redundant database (ClusteredNR) \cite{ncbi2022clusterednr}  using full sequences: all were at least 16 residues distant from the nearest neighbour, with most exceeding 20. Edit-distance distributions are shown in \cref{fig:novelty}.

To avoid selecting closely related variants, we clustered designs by concatenated CDR sequences at \SI{80}{\percent} identity using MMseqs2 clustering with default parameters \cite{steinegger2017mmseqs2}. We advanced a single representative per cluster. A large fraction formed singleton clusters, indicating substantial sequence diversity among candidates.

To increase the likelihood of obtaining biophysically well-behaved, monomeric antibodies, we applied sequence filters excluding designs with cysteines or N-linked glycosylation motifs within designed \gls{cdr} loops.

\paragraph{Experimental methods}
Binder screening and characterization followed a tiered workflow. Initial screening was performed using a single-point \gls{bli} assay to identify candidates with measurable target binding. Designs exceeding a predefined binding response threshold defined in \cref{sec:BLI} were designated as hits and advanced to determine \kd{}.

Preliminary hits were evaluated by 5-point \gls{bli} or \gls{spr}, with final affinities reported only for binders meeting predefined criteria described in \cref{sec:BLI} and \cref{sec:SPR}. Candidates from the initial screen that failed to reproduce binding or yielded poor-quality fits upon follow-up characterization were not considered validated hits.

To investigate binding specificity, targeted point mutations were introduced at selected interface residues across four TNFL9-binding VHH designs. Residues within the CDRH3 loop that engage the same region of the target epitope were identified from the generated all-atom complex and individually mutated to alanine to disrupt key intermolecular contacts.

Developability was assessed using a standard panel of biophysical assays designed to capture key properties relevant to antibody developability, including aggregation propensity, polyreactivity, hydrophobicity, and thermostability. These assays provide complementary measures of protein stability and nonspecific interactions that are commonly associated with clinical attrition risk. All assays were performed alongside therapeutic antibody controls to enable relative benchmarking; full experimental protocols are provided in \cref{sec:developability}.

We assessed aggregation propensity by SEC-HPLC and considered a sample to meet the aggregation criterion if it was at least \SI{90}{\percent} monomeric. For the remaining developability assays, we defined pass criterion relative to therapeutic IgG controls measured alongside our samples. For polyreactivity, we plate-normalized the BVP ELISA signal such that bococizumab was set to $1$ and panitumumab to $0$; samples with a normalized score $\le1$ were considered non-polyreactive. For hydrophobicity, we normalized HIC-HPLC retention times so that lirilumab had a value of $1$, and required samples had a normalized value $\le1$ to meet the criterion. For thermostability, we used the first melting transition ($T_\mathrm{m1}$) and classified samples as thermostable if $T_\mathrm{m1}$ exceeded that of bococizumab which was measured to have a thermostability of \SI{61}{\celsius}.

These therapeutic controls were selected to calibrate each assay against empirically observed ranges for clinical-stage and approved antibodies \cite{Jain2017biophysical}. Bococizumab and panitumumab serve as high and low polyreactivity references to support plate-normalized BVP scoring. Lirilumab provides a high-retention HIC reference, defining an upper bound for acceptable hydrophobicity on our normalized scale. Bococizumab measures on the lower end of permissible thermostability as measured by DSF, making it a permissive reference for classifying variants as thermostable.

\subsection{Immunogenicity assessment}
Four VHH binders targeting TNFL9 were selected from the set of validated hits described in \cref{sec:vhh_scfv_design_methods} as a representative subset for \exvivo{} immunogenicity assessment.

\paragraph{Experimental methods}
\textit{Ex vivo} immunogenicity was assessed using primary human \gls{pbmc} from 10 donors to capture donor-to-donor variability in immune responses. Cytokine release and cellular immune responses were evaluated following exposure to the selected VHH binders across a concentration series, alongside established positive and negative controls. The clinically approved VHH therapeutic caplacizumab was tested at equivalent concentrations as a reference.

Cytokine secretion was quantified from culture supernatants using multiplex Luminex assays, providing a sensitive readout of innate and early adaptive immune signaling. Cellular responses were further assessed using a luminescence-based CellTiter-Glo assay measuring \gls{pbmc} metabolic activity as a surrogate for sustained immune cell viability and expansion. Together, these complementary assays enable detection of both acute inflammatory signaling and longer-term cellular immune responses. Full experimental protocols and assay details are provided in \cref{sec:immunogenicity}.

\subsection{Macrocycle design}
\label{sec:macrocycle_design}
Macrocycle design and filtering followed the computational pipeline described in our prior work \cite{team2025latent}. Macrocycles were sampled across a length range of 6–18 residues to match the typical length ranges of the RaPID mRNA display system, enabling direct comparison \cite{goto2021rapid}. 

Unlike our earlier campaign \cite{team2025latent}, no synthesis feasibility filter was applied. This decision reflected the absence of published evidence that vendor-recommended synthesis rules improved experimental success, and a preference for maximizing sequence diversity and minimizing bias. Designs passing all filters and novelty assessment were ranked by computing the average across all interchain terms in the predicted alignment error matrix (\texttt{ipae}) with the top 10 designs per target selected for experimental characterization.

\paragraph{Experimental methods}
Macrocycle designs were synthesized and cyclized via head-to-tail lactam chemistry to form the peptide bond between the peptide termini. mRNA display control macrocycles were synthesized using an alternative thioether-based cyclization strategy. Details of macrocycle synthesis and cyclization are provided in \cref{app:cyclization}. Binding affinities were measured using an 8-point \gls{spr} assay for high-precision \kd{} determination, see \cref{sec:SPR} for more details.

\section[Discussion]{Discussion}
We have presented \themodel{}, an all-atom generative model that designs antibodies with drug-like properties from the first generation. The model achieves a \SI{50}{\percent} target-level success rate across 18 targets, produces developability profiles matching approved therapeutics, and demonstrates low immunogenicity in human donor panels—addressing a critical gap, given that immunogenicity is a leading cause of clinical failure.

\themodel{} generates antibodies and macrocyclic peptides from a single architecture without task-specific fine-tuning, suggesting the model has learned the biochemistry of specific binding interactions at the atomistic level. The macrocycle results, outperforming exhaustive library screens with only 10 designs, demonstrate that generative precision can replace brute-force screening.

The efficiency gains extend beyond hit finding. By generating molecules with favourable properties from the outset, the model produces candidates closer to clinical leads than conventional starting points. Molecules that clear binding thresholds but fail on developability or immunogenicity often cannot be rescued despite extensive optimization. At the scale of 4 to 24 designs tested, antibody design becomes accessible to individual researchers, and applications such as personalized therapeutics and rapid pandemic response become more feasible.

Drug-like properties emerging without explicit optimization suggests the model has learned principles beyond loop generation. Favorable scaffolds alone do not guarantee success: traditional CDR grafting frequently causes destabilization, aggregation, or loss of binding \cite{jones1986replacing, queen1989humanized, dudgeon2012general}. \themodel{} appears to avoid these failure modes while also producing low polyreactivity and low immunogenicity, suggesting it has learned to generate not only structurally compatible CDRs, but also sequences that avoid promiscuous or immunogenic motifs.

All results are preclinical, with animal studies and clinical trials remaining ahead. Immunogenicity assessment was conducted \exvivo{} using human donor panels—a well established proxy that does not substitute for \invivo{} assessment. Key directions include extending immunogenicity assessment across additional targets, exploring functional assays such as agonism and T-cell engagement, and understanding failure modes on remaining targets. For macrocyclic peptides, future directions include optimizing for cell penetration and oral bioavailability—properties that would unlock intracellular targets currently inaccessible to antibody-format therapeutics. Further work will expand to full IgG and other antibody formats.

\themodel{} is available to selected partners through the Latent Labs Platform at  \href{https://platform.latentlabs.com}{\texttt{\color{latent-purple}platform.latentlabs.com}}.

\section*{Contributors}
Henry Kenlay\textsuperscript{*}, Daniella Pretorius\textsuperscript{*}, Jonathan Crabbé, Alex Bridgland, Sebastian M. Schmon, Agrin Hilmkil, James Vuckovic, Simon Mathis, Tomas Matteson, Rebecca Bartke-Croughan, Amir Motmaen, Robin Rombach\textsuperscript{**}, Mária Vlachynská,
Alexander W. R. Nelson\textsuperscript{**}, David Yuan, Annette Obika, Simon A. A. Kohl\textsuperscript{***}

\textsuperscript{***} Corresponding author. E-mail: \href{mailto:simon@latentlabs.com}{\texttt{simon@latentlabs.com}}.\\
\phantom{*}\textsuperscript{**} Work performed as an advisor to Latent Labs.\\
\phantom{**}\textsuperscript{*} Equal contributions.\\

\paragraph{Author contributions}

\textbf{Conceptualization and team leadership:} S.K. conceived the research direction and priorities for applications, and led the team. D.Y., D.P., led the experimental design with contributions from A.M. A.O., A.N. contributed to project delivery and narrative.

\textbf{Machine learning development:} J.C., S.K developed the model, with contributions from H.K., S.M, S.S. A.B. maintained computational infrastructure for model training. R.R. provided guidance on ML techniques.

\textbf{Computational design and evaluation:} H.K., D.P., S.K. led computational protein design workflows. H.K. constructed computational filters, with contributions from J.C. and T.M. H.K., D.P., A.M. analyzed experimental data. S.S., J.C., H.K. performed computational benchmarking. 

\textbf{Experimental validation:} D.Y. oversaw internal and external validation. R.B.C., A.M. conducted internal experiments and contributed to experimental design. D.Y., A.O., D.P. managed external laboratory partnerships.

\textbf{Model serving via web platform:} A.H. led the platform and inference work. S.K, H.K., A.H. designed the antibody user interface. A.H., J.V., T.M., and S.M. contributed to the Latent-X API and A.H., J.V., S.S., and S.M. contributed to the platform development.

\textbf{Writing and figures:} S.K. oversaw manuscript delivery. S.K., H.K., D.P. wrote the manuscript. H.K., D.P., M.V., A.B., S.M., S.S. made figures. R.B.C., D.Y. contributed to data analysis and manuscript writing. 

All authors contributed to the work and approved the final manuscript.

\paragraph{Competing interests}
All authors have contributed as employees of or advisors to Latent Labs Technologies Inc. or Latent Labs Limited. 

\paragraph{Acknowledgments}
We thank Krishan Bhatt for his help with the organization of our research offsite where key aspects of this work were developed.

\bibliographystyle{unsrtnat}
\bibliography{references}

\clearpage
\appendix
\thispagestyle{empty} %
\section*{\Titlefont Supplementary information}

\renewcommand{\thefigure}{S\arabic{figure}}
\renewcommand{\theHfigure}{S\arabic{figure}}
\setcounter{figure}{0}
\renewcommand{\thetable}{S\arabic{table}}
\setcounter{table}{0}

\section{Model usage}
\label{app:model_inputs}

\themodel{} is available at \href{https://platform.latentlabs.com}{\texttt{\color{latent-purple}https://platform.latentlabs.com}}.  \themodel{} requires both a target specification and a binder specification. The target specification consists of the following: 
\begin{enumerate}
    \item \textbf{Target structure:} The mmCIF file that describes the structure and sequence of the target protein(s) for which binders are designed. The target can consist of multiple protein chains. 
    \item \textbf{Hotspot residues:} The sequence location of the subset of target residues that constitute the target’s binding hotspots. At least one hotspot needs to be provided. In practice, a small number of surface-accessible, spatially close hotspots suffices, and they can be effectively used to steer the model. 
    \item \textbf{Target cropping:} \themodel{} can be conditioned on cropped targets, for example in order to fit the context window. 
\end{enumerate}

Binder specification depends on the design modality (e.g., macrocycles, mini-binders, or antibody-based binders), as detailed below. For macrocycles and mini-binders, a user must specify the binder length. For antibody designs, the user should specify: 
\begin{enumerate}
    \item \textbf{Framework structure:} The mmCIF file that describes the structure and sequence of an antibody scaffold. The scaffold can be specified by a pre-existing antibody, such as a therapeutic Fv or VHH.
    \item \textbf{CDR lengths:} Between each framework region, a CDR loop will be designed by \themodel{}. A user must specify the length of each CDR loop.  
\end{enumerate}

The user should be aware of the following details on input representations:
\begin{itemize}
    \item \textbf{Context length:} The context length for inputs and outputs is 512 residues, counting jointly residues in the target and the binder.
    \item \textbf{Structurally unresolved residues:} Structurally unresolved residues in the target are dropped, which means that the amino acid identities of unresolved residues are not represented in the model's input either.
    \item \textbf{Target backbone only:} Only the backbone atoms of the target structure are provided to the model.
\end{itemize}

\begin{table}
\customtablecaption{\textbf{Sequences and binding affinities of confirmed binders.} Selected binders per target, including \gls{vhh}, \gls{scfv}, and macrocycles. \kd{} values were obtained by \gls{spr} or \gls{bli} (see \cref{sec:experimental_methods}). Reported \kd{} values for controls are replicate values. Macrocycle sequences are denoted by \texttt{cyclo($\cdot$)}.}
\begin{tabular}{
    >{\ttfamily\raggedright\arraybackslash}p{3.2cm}   
    >{\centering\arraybackslash}p{2cm} 
    >{\ttfamily\raggedright\arraybackslash}p{8cm} 
}
\toprule
Binder & \kd{} (M) & Amino acid sequence \\
\midrule
LX2\_scFv\_HDAC8\_10 & $2.62\times10^{-11}$ &
\seqsplit{QVQLQESGPGLVKPSETLSLTCAVSGLSYFKSYGWGWIRQPPGKGLEWIGSITRYGSTYYNPSLKSRVTISVDTSKNQFSLKLSSVTAADTAVYYCARQPIVWSTSDLKGFDVWGQGTLVTVSSGGGGSGGGGSGGGGSGGGGSDIQMTQSPSSLSASVGDRVTITCRASQSVYDYLNWYQQKPGKAPKLLIYRASSLQSGVPSRFSGSGSGTDFTLTISSLQPEDFATYYCQQFNEYPYTFGGGTKVEIK} \\ \addlinespace[0.75ex]

LX2\_scFv\_SAE1\_12 & $1.65\times10^{-7}$ &
\seqsplit{EVQLVESGGGLVQPGGSLRLSCAASGSAFSSYYIHWVRQAPGKGLEWVARIFPYNSSTRYADSVKGRFTISADTSKNTAYLQMNSLRAEDTAVYYCVREEILLDTSTFNLDTAMAYWGQGTLVTVSSGGGGSGGGGSGGGGSGGGGSDIQMTQSPSSLSASVGDRVTITCRASQGVSTSVAWYQQKPGKAPKLLIYSTSFLYSGVPSRFSGSRSGTDFTLTISSLQPEDFATYYCLQYNEYPLTFGQGTKVEIK} \\ \addlinespace[0.75ex]

LX2\_VHH\_1433B\_10 & $2.75\times10^{-9}$ &
\seqsplit{EVQLVESGGGLVQPGGSLRLSCAASQPLSSFYYMGWFRQAPGKGRELVAAISPLTGKTYYPDSVEGRFTISRDNAKRMVYLQMNSLRAEDTAVYYCHASAATPGARTYPASTSDYWGQGTQVTVSS} \\ \addlinespace[0.75ex]

LX2\_VHH\_BHRF1\_3 & $2.70\times10^{-8}$ &
\seqsplit{EVQLVESGGGLVQPGGSLRLSCAASGAWGNIVTMGWFRQAPGKGRELVAAIRMDDGATYYPDSVEGRFTISRDNAKRMVYLQMNSLRAEDTAVYYCMARYAPDRSATVAGVPVVVNKWGQGTQVTVSS} \\ \addlinespace[0.75ex]

LX2\_VHH\_TNFL9\_10 & $3.11\times10^{-8}$ &
\seqsplit{EVQLVESGGGLVQPGGSLRLSCAASLGSSYLAGMGWFRQAPGKGRELVAAIDWATGDTYYPDSVEGRFTISRDNAKRMVYLQMNSLRAEDTAVYYCGAWTTTAAPGTLDPAAVNGFAWGQGTQVTVSS} \\

LX2\_VHH\_LEP\_6 & $4.48\times10^{-8}$ &
\seqsplit{QVQLQESGGGLVQPGGSLRLSCAASPSAGDLYSMAWFRQAPGKERERVAKIYPSSGTTYLADSVKGRFTISQNNAKSTVYLQMNSLKPEDTAMYYCSGLASGSVTVGSTTYNAPKKYWGQGTQVTVSS} \\ \addlinespace[0.75ex]

LX2\_VHH\_MMP2\_7 & $7.70\times10^{-8}$ &
\seqsplit{QVQLQESGGGLVQPGGSLRLSCAASLDWDTVDYMAWFRQAPGKERERVAKIYPSTGKTYLADSVKGRFTISQNNAKSTVYLQMNSLKPEDTAMYYCMANVISVEEVDGKRDFVTKDYWGQGTQVTVSS} \\ \addlinespace[0.75ex]

LX2\_VHH\_AHSP\_9 & $7.97\times10^{-7}$ &
\seqsplit{QVQLQESGGGLVQPGGSLRLSCAASASGYDFVWMAWFRQAPGKERERVAKIFPSDGSTYLADSVKGRFTISQNNAKSTVYLQMNSLKPEDTAMYYCHAVVAKKFEDIKSGDKPPFMFDGWGQGTQVTVSS} \\ \addlinespace[0.75ex]

LX2\_VHH\_ONCM\_3 & $3.26\times10^{-6}$ &
\seqsplit{QVQLQESGGGLVQPGGSLRLSCAASLSGADLLAMAWFRQAPGKERERVAKIYPSFGTTYLADSVKGRFTISQNNAKSTVYLQMNSLKPEDTAMYYCHADALGTQTNEFGFTGFTGLLDGWGQGTQVTVSS} \\ \addlinespace[0.75ex]

LX2\_CYC\_PHD2\_1 & $1.54\times 10^{-9}$ & cyclo(VDHGVYASVNAVTGEY) \\ \addlinespace[0.75ex]

RAPID\_mRNA\_PHD2 & $7.29\times10^{-7}$ & See \cite{mcallister2018non} \\ \addlinespace[0.75ex]

LX2\_CYC\_K-Ras-G12D\_6 & $5.43\times10^{-6}$ & cyclo(FYIPEIDQYLCS) \\ \addlinespace[0.75ex]

RAPID\_mRNA\_K-Ras-G12D & $5.53 \times 10^{-6}$ & See \cite{zhang2020gtp} \\

\bottomrule
\end{tabular}
\label{tab:binder_table}
\end{table}

\section{Computational design}
\cref{tab:target_table} provides details on the target structures we used for computational design. To fit the model's 512 residue context window, we cropped larger targets by removing residues distal to the binding interface. 

In \cref{tab:vhh_frameworks} and \cref{tab:scfv_frameworks} we detail the VHH and scFv frameworks used, respectively, along with the sampled CDRH3 length ranges. Frameworks were annotated using ANARCI with IMGT framework numbering and definitions \cite{dunbar2016anarci}. Note that \texttt{3eak} and \texttt{7ssc} contain an unresolved C-terminal serine at the end of the FW4 region; for wet-lab characterization, we append this residue to the designed sequence. For scFv designs, we modelled binders as Fv fragments during computational design and then added a (G4S)4 linker in the heavy–light orientation after the design step.

\clearpage
\thispagestyle{empty}
\begin{sidewaystable}[p]
\centering
\customtablecaption{\textbf{Protein targets used in experimental validation.} PDB structures \cite{berman2000protein}, disease relevance, and reagent sources are provided.}
\begin{tabular}{
    l
    >{\ttfamily\centering\arraybackslash}p{1.2cm}
    >{\ttfamily\raggedright\arraybackslash}p{5cm}
    >{\arraybackslash}p{4cm}
    >{\arraybackslash}p{2cm}
    >{\ttfamily\raggedright\arraybackslash}p{3cm}
    >{\arraybackslash}p{2cm}
}
\toprule
Target & PDB ID & Residues & Disease areas & Vendor & Catalogue & Valency \\
\midrule
1433B  & 6a5q & B3-B234 & Oncology & Sino Biological& 10843-H09E-B & Bivalent \\
1433E  & 8dgm & A33-A232 & Developmental Neurology & Sino Biological& 50691-M09E & Bivalent \\
AHSP   & 3ia3 & A2-A91 & Haematology & Sino Biological& 14391-HNAE & Monovalent \\
BHRF1  & 2wh6 & A2-A158 & Infectious Disease; Oncology & Cusabio & CSB-EP314488EFC & Monovalent \\
CD98   & 8kdd & A382-A630 & Oncology & Sino Biological& 12206-H07H & Monovalent \\
HDAC8  & 3rqd & A14-A40, A71-A84, A95-A116, A136-A245, A260-A287, A300-A314, A328-A366 & Oncology & Sino Biological& 10864-H09B & Bivalent \\
IL-20  & 4doh & A24-A176 & Autoimmune Disease; Inflammatory Disease & Sino Biological& 13060-HNAE & Monovalent \\
IL-3   & 5uv8 & B15-B30, B33-B121 & Haematology; Oncology & Sino Biological& 11858-HNAE & Monovalent \\
LEP    & 7z3q & A22-A46, A71-A115, A141-A164 & Metabolic Disease & Sino Biological& 10221-HNAE & Monovalent \\
MMP2   & 1gxd & A447-A631 & Oncology & Sino Biological& 10082-HNAH & Monovalent \\
ONCM   & 8v2c & A33-A144, A170-A202 & Oncology & Sino Biological& 50112-MNAE & Monovalent \\
PD-L1  & 5o45 & A17-A132 & Oncology & BioTechne & 9049-B7-100 & Monovalent \\
PRL    & 3npz & A14-A142, A147-A156, A159-A194 & Endocrinology & Sino Biological& 10275-HNAE-B & Monovalent \\
SAE1   & 1y8q & A10-A110, A128-A167, A226-A345 & Oncology & Sino Biological& 13921-HNCB & Monovalent \\
SC2RBD & 6m0j & E333-E526 & Infectious Disease & Sino Biological& 40592-V08B & Monovalent \\
SOMA   & 1hwg & A1-A147, A154-A190 & Endocrinology & Sino Biological& 16122-HNCE & Monovalent \\
TNFL9  & 6a3v & A90-A170, A176-A243 & Immuno-Oncology & Sino Biological& 15693-H07H2 & Trivalent \\
UBE2B  & 2ybf & A3-A152 & Oncology & Sino Biological& 51243-MNCE & Monovalent \\
\kras{} & 6wgn & A1-A34, A38-A169 & Oncology & Sino Biological & R06-32BH & Monovalent \\
PHD2 & 6yw1 & A186-A239, A250-A404 & Haematology & Cusabio & CSB-EP863932HU1b1 & Monovalent \\
\bottomrule
\end{tabular}
\label{tab:target_table}
\end{sidewaystable}
\clearpage

\clearpage
\thispagestyle{empty}
\begin{sidewaystable}
    \centering
    \addtocounter{table}{-1} 
    \begin{subtable}{0.9\textwidth}
        \customtablecaption{\textbf{VHH frameworks and CDRH3 sampling ranges used for computational design.}}
        \centering
        \begin{tabular}{
            @{}       
            >{\ttfamily}l  
            l              
            >{\ttfamily}l  
            >{\ttfamily}l  
            >{\ttfamily}l  
            >{\ttfamily\arraybackslash}l    
        }
        \toprule
        PDB ID & CDRH3 lengths & FWH1 & FWH2 & FWH3 & FWH4 \\
        \midrule
        3eak & 16 - 20 & A1-A25 & A37-A53 & A62-A99 & A118-A127 \\
        5jds & 19 - 23 & B1-B25 & B34-B50 & B59-B96 & B118-B127 \\
        7eow & 19 - 23 & B2-B26 & B35-B51 & B60-B97 & B119-B129 \\
        8hxq & 10 - 14 & A1-A25 & A34-A50 & A59-A96 & A109-A119 \\
        8z8m & \,6 - 10 & B1-B25 & B34-B50 & B59-B96 & B105-B115 \\
        \bottomrule
        \end{tabular}
        \label{tab:vhh_frameworks}
    \end{subtable}

    \vspace{2em}

    \begin{subtable}{0.9\textwidth}
        \customtablecaption{\textbf{Fv framework sequences used for scFv designs and CDRH3 length ranges.}}
        \centering
        \begin{tabular}{
            @{}
            >{\ttfamily}l      
            l                  
            >{\ttfamily}l      
            >{\ttfamily}l      
            >{\ttfamily}l      
            >{\ttfamily}l      
            >{\ttfamily}l      
            >{\ttfamily}l      
            >{\ttfamily}l      
            >{\ttfamily\arraybackslash}l 
        }
        \toprule
        PDB ID & CDRH3 lengths & FWH1 & FWH2 & FWH3 & FWH4 & FWL1 & FWL2 & FWL3 & FWL4 \\
        \midrule
        6zqk & 13 - 17 & A1-A25 & A34-A50 & A59-A96 & A110-A120 & A148-A173 & A180-A196 & A200-A235 & A245-A254 \\
        7kql & 15 - 19 & H1-H25 & H36-H52 & H60-H97 & H113-H123 & L1-L26 & L34-L50 & L54-L89 & L99-L108 \\
        7ssc & 16 - 20 & H1-H25 & H34-H50 & H58-H92 & H103-H127 & L1-L26 & L33-L49 & L53-L88 & L98-L107 \\
        7urx & 15 - 19 & H20-H44 & H53-H69 & H78-H115 & H131-H141 & L20-L45 & L52-L68 & L72-L107 & L117-L126 \\
        7yv1 & 13 - 17 & H1-H25 & H35-H51 & H59-H96 & H110-H120 & L1-L26 & L33-L49 & L53-L88 & L98-L107 \\
        \bottomrule
        \end{tabular}
        \label{tab:scfv_frameworks}
    \end{subtable}
\end{sidewaystable}
\clearpage

\subsection{Sequence novelty of designed antibodies}
\label{app:novelty}
To compute sequence novelty, we compared generated antibody sequences against \gls{sabdab} \cite{dunbar2014sabdab} using concatenated CDR sequences. For each antibody chain (VH or VL) in \gls{sabdab}, we concatenated all corresponding CDR sequences to construct a target database. Sequences generated by \themodel{} were queried against this database using MMseqs2 with default protein settings and the following parameters: \texttt{mmseqs search queryDB targetDB resultDB tmp -s 7.5 --num-iterations 3 -e inf --max-seqs 100000}. For each match, we computed the Levenshtein edit distance over the concatenated CDR sequence, and defined novelty with respect to the minimum edit distance to any database entry.

In addition, sequence similarity searches using the full heavy and light chain sequences were performed with BLASTP \cite{altschul1990basic} against the ClusteredNR database \cite{ncbi2022clusterednr}. 

\begin{figure}[h!]
    \centering
    \includegraphics[width=\textwidth]{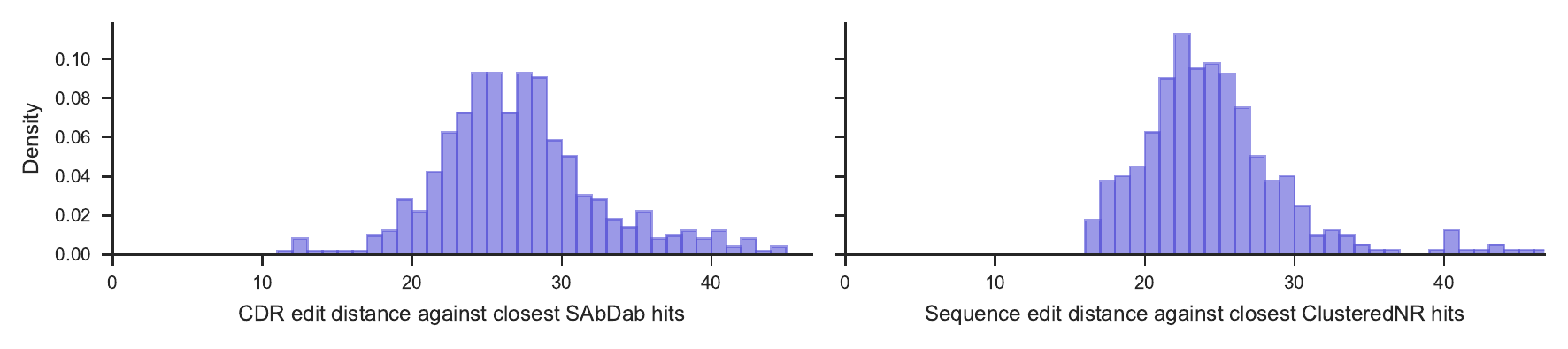}
    \customcaption{\textbf{Sequence novelty of generated antibodies.} Distributions of distances to the closest known antibody sequences, measured as minimum edit distance over concatenated CDR sequences to the nearest SAbDab entry (left) and full-length sequence edit distance to the closest ClusteredNR hit identified by BLASTP (right).}
    \label{fig:novelty}
\end{figure}

\Cref{fig:novelty} shows the distributions of nearest-neighbour distances from both analyses: minimum edit distance over concatenated CDRs to \gls{sabdab} entries and full-length heavy/light-chain similarity to the nearest BLASTP hit in ClusteredNR.

\subsection{Developability benchmark}
\label{sec:developability_benchmark}
We evaluated the developability profile of ten \gls{vhh} and \gls{scfv} domains derived from approved and clinical-stage therapeutics as comparison data points. These domains were extracted from entries in Thera-SAbDab \cite{raybould2020thera} and were taken from entries for rimteravimab, sonelokimab, tarperprumig, brolucizumab, cemavafusp, crefmirlimab, tebentafusp, tarlatamab, and licaminlimab. 

\section{Expected \insilico{} hit rates for VHHs, mini-binders and macrocycles}
\label{app:insilico_hit_rates}

In previous work \cite{team2025latent}, we benchmarked \thepreviousmodel{} on 200 held-out target proteins. For each target, we generated three sets of hotspots and used them to measure the distribution of \insilico{} hit rates. To evaluate the \insilico{} hit rate of \themodel{} for antibody design, we benchmarked the model on the same set of targets and hotspot combinations. We used the five \gls{vhh} frameworks detailed in \cref{tab:vhh_frameworks}, sampling 100 designs per combination of framework, epitope, and target, resulting in $1500$ designs per target. For simplicity, generated \glspl{cdr} match the length of the original VHH from which the framework is extracted.

\begin{figure}[h!]
    \centering
    \includegraphics[width=\textwidth]{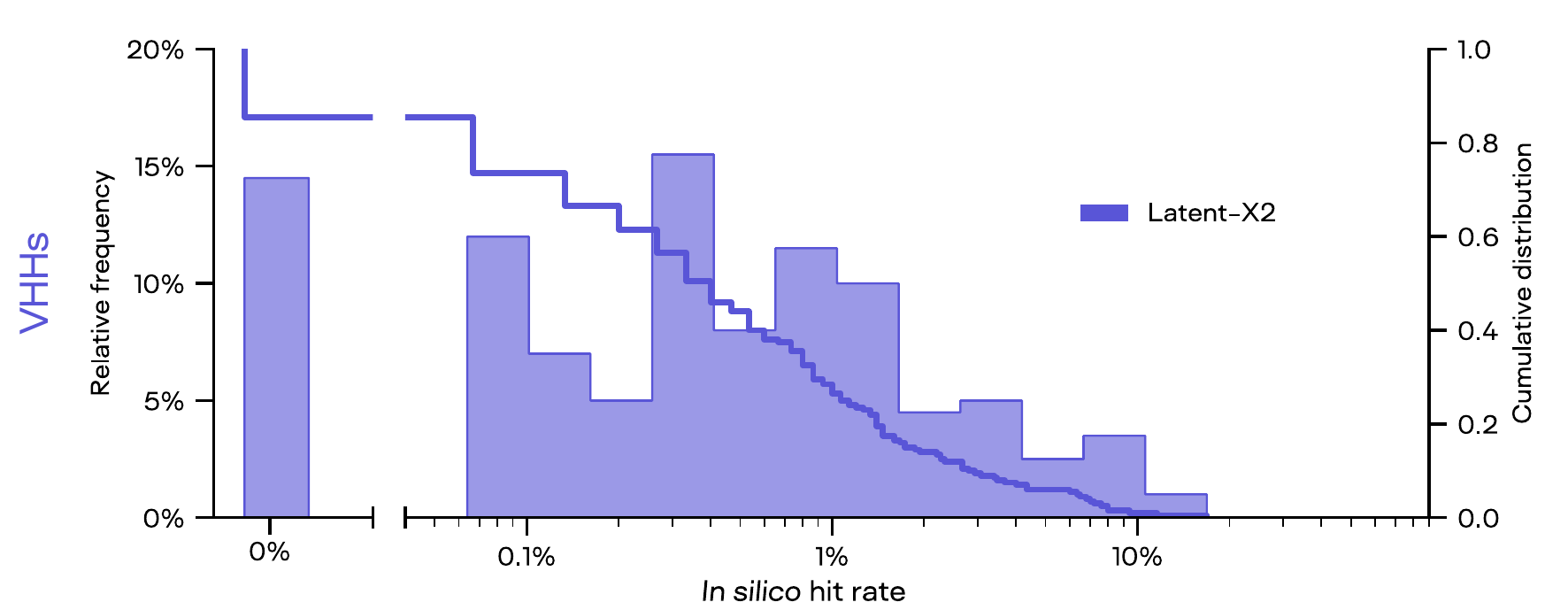}
    \customcaption{\textbf{\Insilico{} hit rates for \themodel{} when designing \glspl{vhh}.} Distribution of \insilico{} hit rates across the held-out benchmark of 200 targets. Distributions show the binned relative frequency against the y-axis on the left; curves show the cumulative distribution functions against the y-axis on the right.}
    \label{fig:dist_plot_vhh}
\end{figure}

The distribution of \insilico{} hit rates is shown in \cref{fig:dist_plot_vhh}. Across the held-out benchmark, most targets yielded at least one passing design, with approximately \SI{85}{\percent} exhibiting a non-zero \insilico{} hit rate. The mean \insilico{} hit rate across targets was \SI{1.15}{\percent}.

We compared the \insilico{} hit rate of \themodel{} with \thepreviousmodel{}, using identical experimental conditions: binder lengths, scoring methodology, $600$ samples per target, and hotspot selection (random but fixed across both models). Both models share the same training data cutoff, ensuring a fair comparison.

\begin{figure}[h!]
    \centering
    \includegraphics[width=\textwidth]{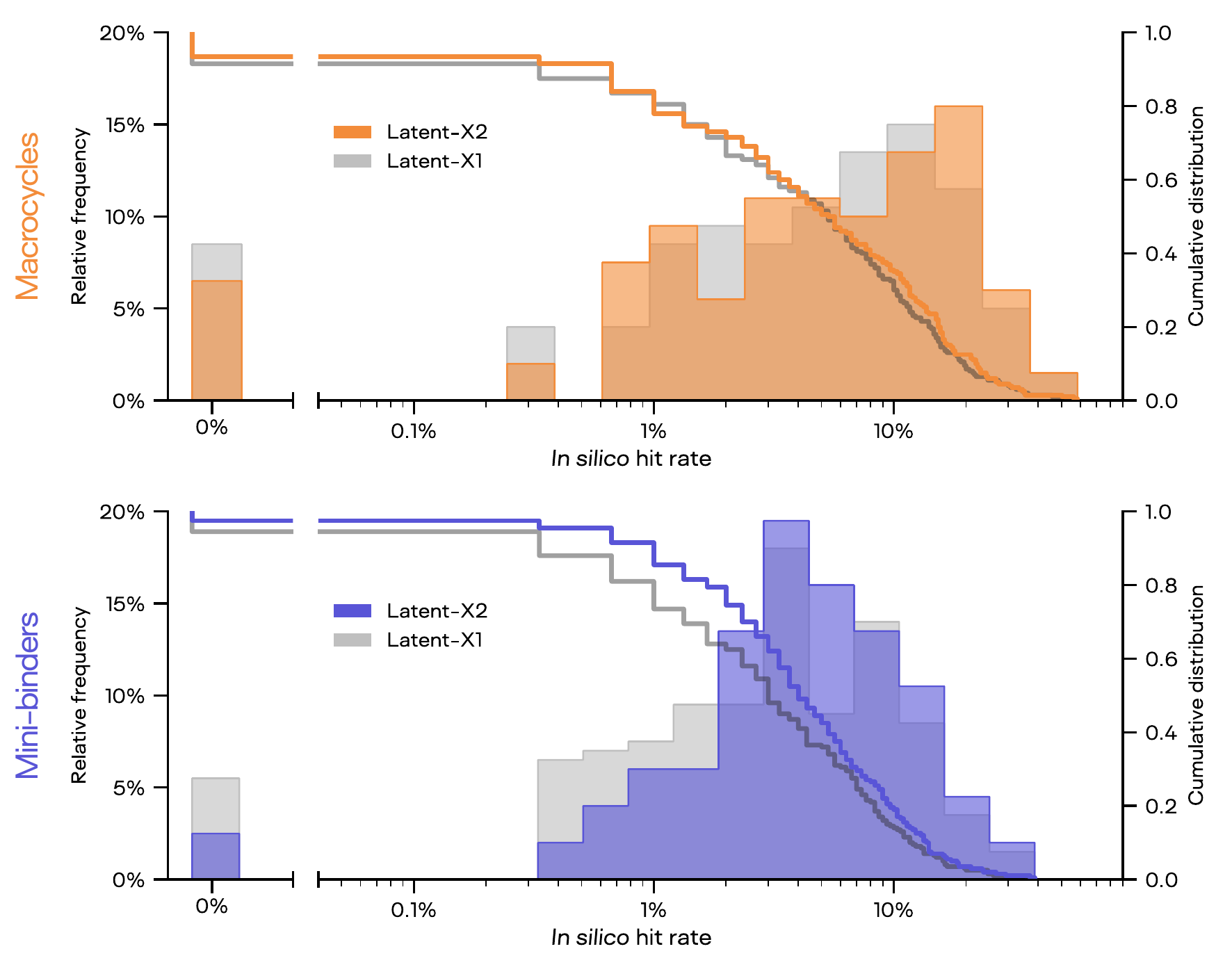}
    \customcaption{\textbf{\Insilico{} hit rates for \themodel{} vs \thepreviousmodel{} across mini-binder and macrocycle modalities.} \themodel{} hit rates are shown in purple and those of \thepreviousmodel{} in grey. \textbf{a)} Distribution of \insilico{} hit rates for mini-binders on 200 held-out targets. \textbf{b)} Distribution of \insilico{} hit rates for macrocycles on 200 held-out targets. Distributions show the binned relative frequency against the y-axis on the left; curves show the cumulative distribution functions against the y-axis on the right.}
    \label{fig:dist_plot_minibinders_and_macrocyles}
\end{figure}

The distribution of \insilico{} hit rates for mini-binders and macrocycles is shown in \cref{fig:dist_plot_minibinders_and_macrocyles}. Quantitatively, the mean success rate for mini-binders increases from \SI{3.92}{\percent} with \thepreviousmodel{} to \SI{6.17}{\percent} with \themodel{}, while macrocycles increase from \SI{5.33}{\percent} to \SI{8.94}{\percent}. Notably, \themodel{} produces passing designs for a larger number of targets, overall reducing the fraction of targets with zero hits. Beyond unlocking new antibody modalities, \themodel{} demonstrates improved \insilico{} performance across mini-binders and macrocycles.

\section{Experimental methods}
\label{sec:experimental_methods}
\subsection{Antibody synthesis and purification}
Antibody binders were produced using either mammalian expression in CHO cells or cell-free \textit{in vitro} transcription–translation. Binders targeting 1433B, 1433E, AHSP, CD98, HDAC8, IL-20, IL-3, LEP, MMP2, ONCM, PRL, SOMA, TNFL9, and UBE2B, as well as the developability benchmark binders, were expressed in CHO cells. Binders targeting BHRF1, PD-L1, SAE1, and SC2RBD were produced using cell-free expression. The ten VHH and scFv benchmark binders were expressed in CHO cells.

For mammalian expression, DNA sequences encoding antibody designs were codon-optimized and synthesized. Sequences were cloned into pcDNA3.4 vectors with a human IgG1 Fc (hinge-CH2-CH3) tag, and plasmids were prepared for transfection into CHO cells. TurboCHO cells were maintained at \SI{37}{\celsius} with \SI{5}{\percent} $CO_2$ on an orbital shaker. Cells were seeded one day prior to transfection. On the day of transfection, DNA and transfection reagent were mixed and added to cells, which were supplemented with feed approximately 24 hours post-transfection.

Antibodies were purified using AmMag™ Protein A Magnetic Beads. After washing, bound antibody was eluted, pooled, and buffer-exchanged into the final formulation buffer. Purified antibodies were analyzed by SDS-PAGE for integrity, SEC-HPLC for molecular weight and purity, and A280 for concentration.

For cell-free expression, antibody designs were codon-optimized and synthesized as full-length double-stranded linear DNA. DNA is then used for cell-free protein expression with a fully reconstituted cell-free system made from individually purified components of the \textit{E. coli} transcription–translation machinery. Expressed proteins (containing a linker, twin-Strep-tag, and a fused split GFP tag) are quantified by the amount of loading detected on the sensors during the BLI/SPR loading cycles as well as a split GFP system using a fused tag.

\subsection{Macrocycle synthesis and cyclization}
\label{app:cyclization}
Peptides were synthesized as follows, using standard Fmoc-based solid-phase peptide synthesis on modified chloride resin. Amino acids were sequentially coupled to the resin with Fmoc deprotection steps using piperidine/DMF. The coupling and deprotection cycle was repeated until the full linear sequence was assembled, with reaction progress monitored by colorimetric resin tests.  Once the linear peptide was fully assembled, it was cleaved from the resin using a 1:3 (v/v) trifluoroethanol/dichloromethane solution, maintaining side-chain protection.

Head-to-tail lactam cyclization was carried out in solution using PyBOP and DIPEA in DMF or DMF/DMSO (2:1 v/v) for two to ten hours, followed by solvent removal via rotary evaporation. For mono thioether–linked cyclization, sequences with a chloroacetylated N-terminus were cyclized following cleavage from the resin to form a thioether linkage. Global deprotection and final cleavage were achieved using TFA-based cocktails for 2--3 hours. The reaction mixture was precipitated into cold tert-butyl methyl ether, centrifuged, and the supernatant decanted to yield the crude macrocycle. The crude product was purified by preparative HPLC with a mobile phase containing \SI{0.1}{\percent} TFA. Collected fractions were analyzed by ESI-MS and analytical HPLC to confirm identity and purity. Fractions with \(>\)\SI{90}{\percent} purity were pooled and lyophilized to obtain final peptide powders.

\subsection{Experimental characterization of binding affinity via SPR}
\label{sec:SPR}
For all macrocycles and antibodies produced in CHO cells \gls{spr} assays were performed using a Biacore 8K system (Cytiva) at \SI{25}{\celsius}. Series S CM5 sensor chips were used for all assays. Target proteins were immobilized by amide coupling using the Amine coupling Kit (Cytiva). Binding measurements were conducted using either multi-cycle kinetics or single-cycle kinetics, depending on the specific interaction characteristics of the analyte and ligand. For each target, either a commercially available IgG or the corresponding natural ligand was included as a target-level positive control to verify assay performance.

Analytes were injected as 8-point or 5-point concentration series, with starting concentrations ranging from \SI{25}{nM} to \SI{250}{\micro M} depending on the sample. Injections were performed at a flow rate of \SI{30}{\micro L\per\minute}. For \kras{} measurements, the association and dissociation times were both \SI{60}{s}. For \gls{phd2} and TNFL9 measurements, the association and dissociation times were \SI{120}{s} and \SI{180}{s}, respectively. Running buffers were selected based on peptide/protein solubility and formulation requirements, these included ultrapure water, DMSO, and PBS buffer.

Data was processed using Biacore 8K Evaluation Software (version 5.0). Sensorgrams were double-referenced using a blank reference surface and buffer-only injections to correct for nonspecific binding and bulk refractive index effects. We used two fitting approaches to determine binding affinity: kinetic fitting and steady-state fitting. The choice of method depended on the interaction characteristics between the analyte and ligand. For interactions with clearly measurable association and dissociation phases, kinetic analysis was performed using multi-cycle kinetics. For fast-on/fast-off interactions, steady-state fitting was used instead, based on equilibrium binding responses.  For kinetic measurements, fits were accepted based on a $\chi^2$ value of less than \SI{10}{\percent} of $R_{\max}$.

For SPR measurements using antibodies produced by cell-free expression, SPR was carried out on a Carterra LSA XT. Cell-free expression products containing intended antibodies were immobilized on the sensor chip surface using fused tags. Target proteins were then injected at 5 concentrations (ranging from \SIrange{3.2}{1000}{nM}). Global fitting of data to a 1:1 model across the whole concentration series was used to determine \kon{}, \koff{}, and \kd{}. Measurements were performed in at least triplicate and included positive controls with expected binding affinities to ensure consistency.

\subsection{Experimental characterization of binding affinity via BLI}
\label{sec:BLI}
For BLI measurements using antibodies produced in CHO cells, assays were performed using a Sartorius Octet RED384 system with corresponding biosensors at \SI{30}{\celsius} and \SI{1000}{\rpm}. Assays were conducted in black polypropylene flat-bottom assay plates (Greiner, \catno5085651) using PBST buffer (\SI{0.03}{\percent} Tween-20 in PBS, pH 7.2). For each target, either a commercially available IgG or the corresponding natural ligand was included as a target-level positive control to verify assay performance.

Prior to running BLI, biosensors were hydrated and conditioned as described above. Antibodies were diluted to appropriate concentrations in PBST for immobilization. Sensors were immersed in the antibody solution and loaded to an approximate \SI{1}{\nano\meter} shift, followed by equilibration in PBST for ~\SI{300}{\second}. Antigen was serially diluted in PBST to suitable concentrations (one or five concentrations). Association was measured by immersing sensors in each antigen concentration solution, and dissociation was monitored by transferring sensors back to PBST. A zero-concentration analyte (PBST buffer alone) was included as a reference.

Single-point BLI assays were used for initial screening, with reference-subtracted binding responses greater than \SI{0.1}{nm} classified as hits and advanced for kinetic characterization.

Data was processed using Octet RED BLI Discovery software version 12.2.2.26. In preliminary experiments, non-specific binding was tested to determine appropriate experimental conditions for subsequent analyses. For multi-point BLI measurements, kinetic parameters were obtained by global fitting across all tested concentrations. Fit quality was assessed using the coefficient of determination (\(R^2\)), calculated over the association and dissociation phases. Kinetic data are reported for fits with \(R^2 > 0.95\). For a single target (AHSP) where no interactions met this criterion, fits with \(R^2 > 0.90\) were included provided they also passed visual quality assessment. For ONCM and LEP, rapid dissociation required shortened dissociation windows (approximately \SI{200}{\second}) to enable robust kinetic fitting.

For BLI measurements using antibodies produced by cell-free expression, assays were performed using a Gator Bio Pro. Antibody constructs were immobilized on the sensor surface via fused affinity tags. Target proteins were injected at four concentrations (\SIrange{31.6}{1000}{nM}), along with a zero-analyte control, and wavelength shifts were recorded at \SI{5}{\hertz}. Global fitting of data to a 1:1 model across the whole concentration series was used to determine \kon{}, \koff{}, and \kd{}. Measurements were performed in at least triplicate and included positive controls with expected binding affinities to ensure consistency.

\subsection{Experimental characterization of antibody developability}
\label{sec:developability}

\paragraph{BVP ELISA}
Baculovirus particles (BVP) contain diverse surface components, including viral proteins, lipids, and host-derived contaminants and were used as antigens in incubation with antibodies in ELISA to evaluate their polyreactivity \cite{Hotzel2012strategy, Jain2017biophysical}. BVPs were coated onto ELISA plates and incubated at low temperature. Subsequent steps, including washing, blocking, sample incubation, secondary antibody binding, and color development, were carried out at room temperature. After stopping the reaction, absorbance was measured at \SI{450}{nm}. The BVP ELISA Score was calculated as the ratio of the sample to blank absorbance to evaluate the tendency of nonspecific binding. Low binding indicates high specificity, while strong binding suggests potential polyreactivity. Bococizumab was used as the
positive control sample and panitumumab as the negative control sample in each experiment \cite{dyson2020beyond, cohenuram2007panitumumab, Jain2017biophysical, Huhtinen2023flpin}.

\paragraph{NanoDSF}
Differential scanning fluorimetry (DSF) measures changes in fluorescence as a protein’s conformation changes, from which protein thermostability can be estimated as melting temperature (Tm). The fluorescence can come from a protein’s own tryptophan, tyrosine, and phenylalanine residues (intrinsic fluorescence). The assessment here was performed using dye-free nanoDSF on the UNcle instrument. Samples were dispensed into UNi tubes and loaded into the instrument for analysis. The instrument applied a controlled temperature ramp while monitoring changes in intrinsic protein fluorescence to determine the antibody's Tm. Bococizumab was used as the positive control sample and sample buffer as the negative control sample in each experiment \cite{Jain2017biophysical}.

\paragraph{HIC-HPLC}
Hydrophobic interaction chromatography separates proteins based on their hydrophobic properties. During the separation process, the high salt concentration in the mobile phase enhances the binding of the protein to the stationary phase through the salting-out effect, and then the salt concentration is reduced to separate proteins with different hydrophobic properties. Lirilumab was used as the positive control sample and tremelimumab as the negative control sample in each experiment \cite{Jain2017HICRetention, Jain2017biophysical}.

\paragraph{SEC-HPLC}

SEC-HPLC was run in an Agilent 1290 UPLC using Phenomenex Biozen \SI{200}{\angstrom} columns (\SI{4.6}{mm} \(\times\) \SI{150}{mm}) with particle size of \SI{1.8}{\micro\meter}. Samples were first filtered using \SI{0.22}{\micro\meter} membranes prior to injection. \SI{3}{\micro g} of filtered samples were injected and run using flow rate of \SI{3.5}{mL\per\minute}, with UV detection (VWD) at \SI{280}{nm} and mobile phase of \SI{0.1}{mol\per L} \ce{Na2SO4} in \SI{0.118}{mol\per L} Phosphate Buffer (pH \(\num{6.7} \pm \num{0.3}\)). 

\subsection{Ex Vivo Immunogenicity Characterization}
\label{sec:immunogenicity}

\paragraph{PBMC cell thawing and seeding}
\Glspl{pbmc} (StemCell Technologies, \catno 70025.3; Hu PBMNC, Cryo, \SI{5e7}{}~cells/vial) were thawed quickly in a \SI{37}{\celsius} water bath by swirling each vial. Characteristics of donors can be found in \cref{tab:pbmc-donors}. Vials were removed when a small amount of ice remained. The entire vial content was then transferred by pipette to a \SI{50}{mL} conical tube. The vial was rinsed with \SI{1}{mL} of pre-warmed media and added dropwise to the cells while swirling the \SI{50}{mL} conical tube. Cells were washed by adding \SIrange{15}{20}{mL} media dropwise while swirling the tube. Then, cells were centrifuged at 300~\texttimes~\text{g} for \SI{10}{min} at room temperature. The supernatant was carefully removed with a pipette, leaving a small amount of medium to ensure the cell pellet was not disturbed. The cell pellet was resuspended by flicking the tube, after which \SIrange{15}{20}{mL} media was added to the tube. Cell suspension was centrifuged again at 300~\texttimes~\text{g} for \SI{10}{min} at room temperature. Supernatant was carefully removed with a pipette, leaving a small amount of medium to ensure the cell pellet is not disturbed. The cell pellet was resuspended, the suspended cells were then counted using a ViCell automated cell counter. Cells were then seeded in 96-well plates at \num{150000} cells/well in \SI{180}{\micro L} ImmunoCult media (StemCell Technologies, \catno 10971) + \SI{10}{\percent} HI FBS. Two 96-well plates were seeded per \gls{pbmc} donor, one plate for \SI{48}{h} supernatant harvest followed by CellTiter-Glo and another plate for \SI{120}{h} supernatant harvest followed by CellTiter-Glo.

\clearpage
\thispagestyle{empty}
\begin{sidewaystable}[p]
\centering
\customtablecaption{\textbf{Characteristics of \gls{pbmc} donors used in \exvivo{} immunogenicity assays.}}
\begin{tabular}{
    >{\ttfamily\centering\arraybackslash}p{0.7cm}    
    >{\centering\arraybackslash}p{0.7cm}             
    >{\ttfamily\centering\arraybackslash}p{1.3cm}    
    >{\ttfamily\centering\arraybackslash}p{0.7cm}    
    >{\centering\arraybackslash}p{1.1cm}             
    >{\centering\arraybackslash}p{1.cm}              
    >{\centering\arraybackslash}p{1.2cm}             
    >{\ttfamily\centering\arraybackslash}p{0.9cm}    
    >{\ttfamily\centering\arraybackslash}p{1.1cm}    
    >{\ttfamily\raggedright\arraybackslash}p{2.7cm}  
    >{\ttfamily\raggedright\arraybackslash}p{2.7cm}  
    >{\ttfamily\raggedright\arraybackslash}p{2.7cm}  
}
\toprule
Gender & Age & Ethnicity & Smoker & Weight (kg) & Height (cm) & Viability (\si{\percent}) & Blood Type & CMV Status & HLA-A & HLA-B & HLA-C \\
\midrule
M & 57 & WHT & Y & 90  & 176 & 96   & A+ & UNK  & A*01:01, A*25:01 & B*08:01, B*18:01 & C*07:01, C*12:03 \\
M & 29 & CAU & N & 83  & 172 & 99   & O+ & NEG & A*01:01, A*02:01 & B*08:01, B*40:01 & C*03:04, C*07:01 \\
M & 42 & BLK & N & 96  & 183 & 96   & A+ & NEG & A*30:02, A*30:01 & B*18:01, B*42:01 & C*07:01, C*17:01 \\
M & 24 & MIX & N & 88  & 181 & 95   & O+ & POS & A*11:01, A*24:02 & B*15:02, B*51:01 & C*08:01, C*15:02 \\
F & 50 & BLK & N & 100 & 155 & 97   & O+ & NEG & A*02:03, A*25:01 & B*08:01, B*13:01 & C*04:03, C*07:01 \\
M & 57 & WHT & Y & 90  & 185 & 98   & A+ & NEG & A*02:01, A*02:01 & B*08:01, B*27:05 & C*02:02, C*07:01 \\
F & 21 & HSP & N & 57  & 168 & 99   & O+ & POS & A*02:01, A*24:02 & B*39:06, B*50:01 & C*04:01, C*07:02 \\
F & 43 & AFA & N & 73  & 163 & 99.9 & A+ & POS & A*03:01, A*68:01 & B*41:02, B*58:02 & C*06:02, C*17:01 \\
M & 20 & ASN & N & 59  & 163 & 99   & B+ & POS & A*33:03, A*68:01 & B*07:02, B*58:01 & C*03:02, C*07:02 \\
F & 32 & CAU & N & 56  & 170 & 99.8 & A+ & POS & A*03:01, A*25:01 & B*40:01, B*57:01 & C*03:04, C*03:04 \\
\bottomrule
\end{tabular}
\label{tab:pbmc-donors}
\end{sidewaystable}
\clearpage

\paragraph{Treatment}
Five treatment antibodies were added at 10\texttimes\ to each donor set of \gls{pbmc} plates (\SI{20}{\micro L} of 10\texttimes\ antibody to \SI{180}{\micro L} of seeded cells) in a four-point, three-fold dilution series run in duplicate, with a top concentration of \SI{30}{\micro g\per mL}. The controls were: 1) Untreated; 2) Positive Control 1: ImmunoCult Human CD3/CD28 T-cell Activator (\SI{25}{\micro L\per mL} treatment) + IL-2 (\SI{10}{ng\per mL}); and 3) Positive Control 2: Invitrogen\texttrademark~eBioscience\texttrademark~Phytohemagglutinin-L (PHA-L) Solution (500\texttimes, \SI{1.25}{mg\per mL}), final treatment concentration \SI{10}{\micro g\per mL}.

\begin{table}[h]
    \centering
    \customtablecaption{\textbf{Positive controls used in \exvivo{} immunogenicity assays.}}
    \begin{tabular}{p{6.5cm} >{\ttfamily\raggedright\arraybackslash}p{2cm} p{4cm}}
    \toprule
    Item & Catalogue & Supplier \\
    \midrule
    ImmunoCult Human CD3/CD28 T-cell Activator & 10971 & StemCell Technologies \\
    Human Recombinant IL-2, ACF & 78145 & StemCell Technologies \\
    Phytohemagglutinin-L (PHA-L) Solution & 50-112-9264 & Invitrogen™ eBioscience™ \\
    \bottomrule
    \end{tabular}
    \label{tab:immuno_controls}
\end{table}

\paragraph{Supernatant harvest for Luminex and CellTiter-Glo assays at 48 hours and 120 hours}
Thirty minutes prior to timepoint, plates were pulled from incubator and placed on bench at RT. Plates were centrifuged at 300~\texttimes~\text{g} for \SI{1}{min}. Using a 12 channel P200, \SI{100}{\micro L} of media were pipetted from the top of the well without disturbing cells at the bottom and transferred to a clear U bottom polypropylene 96 well plate (Corning, \catno 3359). Plate was sealed with foil sticker and transferred to \SI{-80}{\celsius} for Luminex assay. Equal volume of CTG was added to each assay well. Plate was then gently shaken for \SI{2}{min} at \SI{300}{rpm} (covered with foil) then let sit at RT covered for \SI{10}{min}. Luminescence was then read on SpectraMAX.

\paragraph{Luminex assay 120-hour supernatant timepoint}
Supernatant plates were thawed on ice. Six ProcartaPlex Assay plates were run according to the ProcartaPlex Human Mix and Match Panels User Guide Publication Number MAN0024966. Kit components are listed in \cref{tab:kit_components_luminex}. Supernatant \SI{120}{h} sample was diluted 1:2 using assay medium (ImmunoCult-XF T Cell Exp Medium; StemCell Technologies, \catno 10981) + \SI{10}{\percent} HI FBS as diluent. The controls were: 1) Untreated; and 2) Positive control: ImmunoCult Human CD3/CD28 T-cell Activator + IL-2. Plates were read on Luminex FLEXMAP 3D instrument for the following targets: IFN-$\gamma$, IL-10, IL-16, IL-6, IL-8 (CXCL8), and TNF-$\alpha$; see \cref{tab:6plex_standards} for target details and standard concentrations. 

Raw data files were exported from Luminex FLEXMAP 3D and then uploaded to ThermoFisher's ProcartaPlex Analysis App, where data was analyzed and quantified into \si{pg\per mL} per analyte according to the standard curve. When both technical replicates yielded valid measurements, values were averaged prior to downstream analysis. Measurements falling outside the quantifiable range of the standard curve were flagged as out-of-range (OOR). If one replicate yielded a quantifiable value within the standard curve while the other was OOR, the quantifiable value was used for downstream analysis. If both replicates were below the lower limit of quantification (LLOQ), the reported concentration was set to 0. If both replicates exceeded the upper limit of quantification (ULOQ), the reported concentration was designated as above range. This approach allowed retention of samples with partial quantifiability while avoiding extrapolation beyond the validated assay range.

\begin{table}[h]
    \centering
    \customtablecaption{\textbf{Luminex kit components used in cytokine release assays.}}
    \begin{tabular}{p{6cm} >{\ttfamily\raggedright\arraybackslash}p{3cm} >{\ttfamily\raggedright\arraybackslash}p{3cm}}
    \toprule
    Item & Catalogue & Lot \\
    \midrule
    ProcartaPlex Human Mix \& Match Panel & PPX-06-MXPRNXR & 470023-000 \\
    Standard Mix A & S10007EX & 445156-000 \\
    Standard Mix B & S10011EX & 420280-000 \\
    Standard Mix E & S10014EX & 397396-000 \\
    6-Plex Capture Beads & B-06-MXPRNXR-EX & 469917-000 \\
    6-Plex Detection AB & BK-06-MXPRNXR-EX & 469918-000 \\
    Streptavidin-PE (SA-PE) & SA-PE & 428706-000 \\
    Wash Buffer (10X) & WBEX/28 & 25097135 \\
    \bottomrule
    \end{tabular}
    \label{tab:kit_components_luminex}
\end{table}

\begin{table}[h]
    \customtablecaption{\textbf{Cytokine analytes and standard concentrations for Luminex 6-plex assay.}}
    \centering
    \begin{tabular}{p{3cm} p{3cm} p{3cm} p{3cm}}
    \toprule
    Target Name & Bead Number & Concentration (\si{pg\per mL}) & Standard Mix \\
    \midrule
    IFN-$\gamma$ & 43 & 44200 & Standard Mix A \\
    IL-10 & 28 & 4700 & Standard Mix A \\
    IL-16 & 47 & 29600 & Standard Mix E \\
    IL-6 & 25 & 46800 & Standard Mix A \\
    IL-8 (CXCL8) & 27 & 9600 & Standard Mix B \\
    TNF-$\alpha$ & 45 & 29500 & Standard Mix A \\
    \bottomrule
    \end{tabular}
    \label{tab:6plex_standards}
\end{table}

\end{document}